\documentclass[11pt,a4paper,numbers,sort&compress]{article}
\usepackage[utf8]{inputenc}
\pdfoutput=1
\usepackage{jheppub}
\usepackage{graphicx}
\usepackage{amsthm}
\usepackage{mathtools}
\usepackage{empheq}
\usepackage{bbm}
\usepackage{tikz}
\usetikzlibrary{matrix,arrows,shapes,decorations.fractals,decorations.pathmorphing,shadows,patterns,decorations.markings}
\usepackage{dsfont}
\usepackage{slashed}
\usepackage{esvect}

\definecolor{yellow}{cmyk}{0,0,1,0}
\definecolor{magenta2}{cmyk}{0.25,0.8,0,0.1}
\definecolor{cyan2}{cmyk}{0.9,0.02,0,0.1}
\definecolor{green2}{cmyk}{1,0,1,0.15}
\definecolor{oran}{cmyk}{0,0.2,0.57,0}

\newcommand{\cvv}[1]{\reflectbox{\ensuremath{\vv{\reflectbox{\ensuremath{#1}}}}}}
\def\be{\begin{equation}}
	\def\ee{\end{equation}}
\def\bea{\begin{eqnarray}}
\def\eea{\end{eqnarray}}

\newcommand{\rmd}{\,\mathrm{d}}
\newcommand{\zerovec}{\vec{0}}

\newcommand{\flati}[1]{{\underline{#1}}}

\newcommand{\e}[1]{\operatorname{e}^{#1}}

\providecommand{\sgn}{\operatorname{sgn}}
\newcommand\PP[2][0]{\setlength\shlength{#1pt}%
  \stackengine{-5.6pt}{$#2$}{\smash{$\kern\shlength%
    \stackengine{7.55pt}{$\mathchar"017E$}%
      {\rule{\widthof{$#2$}}{.57pt}\kern.4pt}{O}{r}{F}{F}{L}\kern-\shlength$}}%
      {O}{c}{F}{T}{S}}

\newcommand{\PU}{\vv{U}}
\newcommand{\AU}{\cvv{U}}

\preprint{LCTP-23-02}

\title{Beyond AdS$_2$/dCFT$_1$: Insertions in Two Wilson Loops}

\author[a]{Diego H. Correa,}
\author[b,c]{Alberto Faraggi,}
\author[d,e]{Wolfgang M\"uck,}
\author[f,g,h]{Leopoldo A. Pando Zayas,}
\author[a]{Guillermo A. Silva}

\emailAdd{correa@fisica.unlp.edu.ar,
afaraggi@uc.cl, mueck@na.infn.it, lpandoz@umich.edu, silva@fisica.unlp.edu.ar}

\affiliation[a]{Instituto de F\'isica La Plata - CONICET \& Departamento de F\'isica,\\ Universidad Nacional de La Plata,
C.C. 67, 1900, La Plata, Argentina}
\affiliation[b]{Departamento de Ciencias F\'isicas, Facultad de Ciencias Exactas, Universidad Andr\'es Bello,\\
Sazi\'e 2212, Piso 7, Santiago, Chile.}

\affiliation[c]{Instituto de F\'isica, Pontificia Universidad Cat\'olica de Chile,\\
Av. Vicu\~na Mackenna 4860, Santiago, Chile.}

\affiliation[d]{Dipartimento di Fisica “Ettore Pancini”, Universit\`a degli Studi di Napoli ``Federico II''\\
Via Cintia, 80126 Napoli, Italy}
\affiliation[e]{Istituto Nazionale di Fisica Nucleare, Sezione di Napoli\\
Via Cintia, 80126 Napoli, Italy}
\affiliation[f]{Leinweber Center for Theoretical Physics, 
University of Michigan, Ann Arbor, MI 48109, USA}
\affiliation[g]{School of Natural Sciences, Institute for Advanced Study, Princeton, NJ 08540, USA}
\affiliation[h]{The Abdus Salam International Centre for Theoretical Physics, 34014 Trieste, Italy}

\abstract{We consider two-point correlators of local operator insertions in a system of two Wilson-Maldacena loops in ${\cal N}=4$ supersymmetric Yang-Mills theory on both sides of the AdS/CFT correspondence.  On the holographic side the correlator of two Wilson-Maldacena loops is given by a classical string world-sheet which in one phase connects two asymptotically AdS$_2$ regions and in the other phase is given by two disconnected AdS$_2$ caps; this configuration breaks supersymmetry as well as conformal invariance. We present a complete systematic account of the string world-sheet fluctuations, including the fermionic sector, and study the behavior of the holographic two-point correlators. On the field theory side we compute certain two-point correlators of local operator insertions by resumming sets of ladder diagrams. Our results demonstrate the efficacy of previously developed methods in tackling this non-conformal, non-susy regime.}

\date{\today}

\begin{document}

\maketitle


\newpage
\section{Introduction}
The duality between ${\cal N}=4$ supersymmetric Yang-Mills and strings in AdS$_5\times$ S$^5$ is the paradigmatic example of the AdS/CFT correspondence \cite{Maldacena:1997re}, and the Wilson-Maldacena half-BPS operators have played a central role in it since the very inception of the conjecture \cite{Maldacena:1998im,Rey:1998ik,Drukker:2000rr,Erickson:2000af,Drukker:2000ep}. 

The drive for increasingly precise computations of vacuum expectation values on each side of the correspondence has proven to be quite a fruitful direction. Pestun's localization result \cite{Pestun:2007rz} proving the conjecture that expectation values of such half-BPS Wilson loops  are given by the Gaussian matrix model opened the door to various precision tests. In particular, given that the holographic side describes the Wilson loop as a semi-classical string, precision studies beyond the leading order helped to understand aspects of string perturbation theory in this background \cite{Kruczenski:2008zk,Kristjansen:2012nz,Forini:2015bgo,Faraggi:2016ekd,Cagnazzo:2017sny,Medina-Rincon:2018wjs} and paved the way for successful holographic tests based on string theory on other backgrounds \cite{Medina-Rincon:2019bcc,David:2019lhr}.

Since the classical string world-sheet has an AdS$_2$ geometry and the Wilson-Maldacena loop can be interpreted as a defect CFT$_1$ inside ${\cal N}=4$ SYM, this setup furnishes a rigorous instance of AdS$_2$/CFT$_1$ descending directly from string theory on AdS$_5\times S^5$. In this context, some quite impressive results for two-point  \cite{Cooke:2017qgm} and four-point \cite{Giombi:2017cqn} correlators  of certain protected operators have been obtained. The results are in complete agreement with the expectations of CFT$_1$, and highly non-trivial information about anomalous dimensions as a function of the coupling can be read off from them. Moreover, these results are corroborated by other independent methods including the integrability-based quantum spectral curve method \cite{Grabner:2020nis,Julius:2021uka,Cavaglia:2021bnz} and analytic bootstrap methods \cite{Ferrero:2021bsb}.

In this manuscript we build on these interesting developments and initiate the study of correlators between insertions of operators in a system of two Wilson-Maldacena loops \cite{Correa:2018lyl}. A system of two Wilson-Maldacena loops provides a setup where both conformal symmetry as well as supersymmetry are broken. The holographic dual of such a system was first discussed in \cite{Zarembo:1999bu,Olesen:2000ji}. It is encouraging that even in this context we are able to make progress and compute, in various approximations, the two-point correlators on either side of the correspondence following relatively standard techniques. 

We are also motivated by the fact that the expectation value of two Wilson-Maldacena loops is described, in one phase,  by a string world-sheet that connects two asymptotically AdS$_2$ regions where the loops are placed, a Euclidean wormhole. On the other phase, after the Gross-Ooguri transition \cite{Gross:1998gk}, the configuration dominating the partition functions is given by two disconnected AdS$_2$ world-sheets. This setup comes, in some aspects, tantalizingly close to a framework in which important developments in nAdS$_2$/nCFT$_1$ have recently taken place to clarify aspects of Hawking radiation (see \cite{Almheiri:2020cfm} for a review). We hope to eventually connect to such a framework, some encouraging new evidence has recently been reported in \cite{Giombi:2022pas}.

The rest of the manuscript is organized as follows. In section \ref{Sec:GravitySolution} we review the classical string world-sheet that is holographically dual to a configuration of  two Wilson-Maldacena loops. Section \ref{Sec:StringFluctuations} presents a detailed account of the string fluctuations; we obtain the equations of motions for the bosonic and fermionic fields. In section \ref{Sec:Correlators} we discuss the holographic computation of the correlators. In section \ref{Sec:FTside} we describe the field theoretic problem, introduce a number of technical resources needed to sum certain sets of diagrams and present explicit expressions for the correlators of insertions in a system of two Wilson-Maldacena loops. We conclude in section \ref{Sec:Conclusions} where we also point to some interesting open questions that our work motivates. We relegate some technical details to a couple of appendices.

\section{Two holographic Wilson-Maldacena loops}\label{Sec:GravitySolution}

A locally supersymmetric Wilson-Maldacena loop in $\mathcal{N}=4$ SYM theory is given by 
\begin{equation}
 W(C;n^I)= {\rm tr}\, P\exp \left(  g_{_{  Y\!M}}
 \oint_C \rmd t\,  \left(i A_\mu\dot{x}^\mu  + \Phi_I n^I |\dot{x}|\right)\right)  ~.
 \label{WLdef}
\end{equation}
Here, both the gauge field $A_\mu$ and the scalars $\Phi_I$ ($I=1,2,\ldots,6$) are assumed to be in the fundamental representation of the gauge group $U(N)$. The coupling to $\Phi_I$, with $n^I$ being a 6-d unit vector, was introduced by Maldacena \cite{Maldacena:1998im} and is crucial for supersymmetry. 

In this manuscript, we consider two loops, with coaxial circular contours, $C_1$ and $C_2$, of opposite orientations and scalar couplings $n^I_1$ and $n^I_2$ \cite{Correa:2018lyl}, 
\begin{equation}
    \label{C12}
\begin{aligned}
C_1: \quad x^\mu_1(\phi) &= (R_1 \cos \phi, R_1\sin \phi,0,0)~,
\qquad & n^I_1 &=(0,0,0,0,0,1)~,
\\
C_2: \quad  x^\mu_2(\phi) &= (R_2 \cos \phi, -R_2\sin \phi,h,0)~,
 & n^I_2 &=(0,0,0,0,\sin\gamma,\cos\gamma)~.
\end{aligned}
\end{equation}
This configuration, with the contours placed on two parallel planes separated by a distance $h$, can be shown to be equivalent, by a conformal transformation, to a configuration of two concentric loops on the same plane. In appendix \ref{App:ConfTra} we review, for the benefit of the reader, the conformal transformations relating them and show that they are all characterized by the invariant parameter 
\begin{equation}
    \label{conf.invar}
    \alpha = \frac{2R_1 R_2}{h^2+R_1^2 +R_2^2}~.
\end{equation}
Thereby, the approach to be discussed below provides a unified description for various configurations considered in the literature, for example those in the original papers \cite{Zarembo:1999bu,Olesen:2000ji}, as well as configurations obtained using integrability \cite{Burrington:2010yb}. We remind the reader that the setup of separated parallel loops was instrumental to describe the Gross-Ooguri phase transition \cite{Gross:1998gk}. 

In this section, we re-derive the background solution. Second, we characterize its geometry, which we will need for the field equations of the fluctuations in section~\ref{Sec:StringFluctuations}. Last, we repeat the calculation of the on-shell effective action with a renormalization different from \cite{Correa:2018lyl} and show that the results remain unchanged. This implies that the phase transition between the connected and disconnected configurations is a robust  phenomenon.

\begin{figure}[ht]
\center{
\begin{tikzpicture}
\draw  [red, thick]  (1,1) circle (1) ;
\draw [ ->, red, thick] (2,1) arc (0:30:1);
\draw [blue, thick] (1,1) circle (2);
\draw [<-, blue, thick] (3,1) arc (0:30:2);
\draw[->,line width=.02cm,red] (1,1) - - (1- 1*cos 30 , 1-1*sin 30) node[below]{$R_1$};
\draw[->,line width=.02cm,blue] (1,1) - - (1-2*cos{(120)} ,1-2*sin{(120)}) node[below]{$R_2$};
\node (B) at (1+1*cos{(120)},1+1*sin{(120)}) {};
\node (C) at (1+3*cos 30,1+3*sin 30) {}; 
\node (D) at (1+3*cos{(120)} ,1+3*sin{(120)}) {};
\end{tikzpicture}}
\caption{Two concentric Wilson loops with their respective orientations.}
\end{figure}
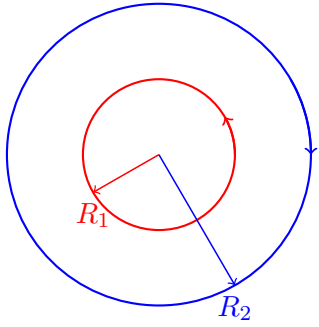

\subsection{Background solution}

Let us start by considering the AdS$_5\times S^5$ bulk solution. Its metric is
\begin{equation}
\label{bg:bulk.metric}
	\rmd s^2 = \frac{L^2}{z^2} \left( \rmd z^2 +\rmd r^2 +r^2 \rmd\phi^2+\rmd x^2 -\rmd t^2 \right) +L^2 \left( \rmd\theta^2 + \sin^2\theta \rmd\Omega_4^2\right).
\end{equation}
Below $\vec{x}=(x,t)$ will denote a Lorentzian 2-vector. In addition, the background geometry is supported by a self dual five-form field strength, 
\begin{equation}
\label{ferm:F5}
	F_5 = \frac4{L} \left(1+\ast \right) \epsilon_{S^5}~,
\end{equation}
where $\epsilon_{S^5}$ is the volume form of the $S^5$ part of the bulk (with radius $L$, not the unit $S^5$). $F_5$ will be relevant only for the fermion fields.

Exploiting the conformal invariance, we consider the string configuration in the form  in which the two boundaries lie on concentric circles on the same plane, $x=t=0$, and have radii $R_1=R_-$ and $R_2=R_+$. For the time being, we may take $R_+\geq R_-$ without loss of generality. Analyzing the world-sheet using concentric rings has the advantage that there is one less variable to consider, namely $x$, compared to the general set-up. Moreover, the $x$-direction becomes a normal direction and is manifestly on the same footing as the $t$-direction. This is helpful when parameterizing the fluctuations.

The world-sheet has a cylinder topology with coordinates $(\tau,\sigma)$, where $\sigma\sim\sigma+2\pi$. Our ansatz for the string embedding is $z=z(\tau)$, $r=r(\tau)$, $\theta=\theta(\tau)$, $\phi=\sigma$, while $x$, $t$ and the position on $\Omega_4$ are fixed. Thus, the induced metric reads ($'=\frac{\rmd}{\rmd\tau}$)
\begin{equation}
\label{bg:induced.metric}
	g_{\alpha\beta} = 
		\begin{pmatrix}
		  \frac{L^2}{z^2} \left(z'^2+r'^2+z^2 \theta'^2\right) &0 \\
		  0 & \frac{L^2r^2}{z^2} 		  
		\end{pmatrix}~.
\end{equation}
With our ansatz, the Euclidean Nambu-Goto action takes the form 
\begin{equation}
\label{bg:NG.action}
	S_{NG} = \sqrt{\lambda} \int \rmd\tau \frac{r}{z^2} \sqrt{z'^2+r'^2+z^2 \theta'^2}~,
\end{equation}
with $\lambda = L^4/\alpha'^2$. Notice we still have the freedom to fix $\tau$-reparameterizations (see \cite{Dekel:2013kwa} for a similar approach).

The solution we are after can be characterized in terms of two conserved charges. The rigid symmetries of \eqref{bg:NG.action} are $\theta\to \theta+\delta\theta$ and the scaling symmetry $r \to c r$, $z\to cz$. The associated Noether charges are 
\begin{equation}
\label{bg:K.def}
	K = \frac{r\theta'}{\sqrt{z'^2+r'^2+z^2 \theta'^2}}~,
\end{equation}
and
\begin{equation}
\label{bg:C.def}
	C = \frac{r(zz'+rr')}{z^2\sqrt{z'^2+r'^2+z^2\theta'^2}}~,
\end{equation}
respectively.

To proceed, we solve \eqref{bg:K.def} for $\theta'$ (without loss of generality we can assume $\theta'\geq0$) and substitute the solution into \eqref{bg:C.def}. It is, however, useful to change variables by setting
\begin{equation}
\label{bg:rho.psi.def}
	z = \rho \sin\psi~,\qquad r = \rho \cos\psi~,
\end{equation}
so that the two world-sheet boundaries are given by $\psi=0$ and $\rho=R_\pm$. The entire procedure gives rise to the differential equation
\begin{equation}
\label{bg:rho.eq1}
	\rho'^2 \left( \cos^2 \psi - K^2 \sin^2 \psi -C^2 \sin^4 \psi \right) = C^2 \rho^2 \psi'^2 \sin^4\psi ~.
\end{equation}
For $C\neq0$, we can take $\psi$ to be a function of $\tau$ which increases from zero to a certain maximum at $\tau=\tau_0$ and then decreases back to zero at the other boundary.\footnote{The solution describing a single Wilson loop (with only one boundary) has the parameters $C=K=0$. Then, $\rho=R$ and  $\psi=\tau \in [0,\frac{\pi}2]$. Other solutions with $C=0$ can be obtained by setting $C\to 0$ in the equations below.} From \eqref{bg:rho.eq1}, the maximum $\psi_0=\psi(\tau_0)$ satisfies 
\begin{equation}
\label{bg:K.sub}
	K^2 = \cot^2 \psi_0 - C^2 \sin^2 \psi_0~,
\end{equation}
which we may use to eliminate $K$ or $C$ in favour of $\psi_0$. 
After eliminating $K$, \eqref{bg:rho.eq1} gives rise to 
\begin{equation}
\label{bg:rho.eq2}
	\frac1\rho \rho' = \frac{C\sin\psi_0\sin^2\psi |\psi'|}{\sqrt{(\sin^2\psi_0-\sin^2\psi)(1+C^2\sin^2\psi_0\sin^2\psi)}}~.
\end{equation}
Integrating this along the entire string yields
\begin{equation}
\label{bg:R.frac}
	\ln \frac{R_+}{R_-} = 2 J~,
\end{equation}
where we have abbreviated\footnote{The integral $J$ was called $F(s,t)$ in \cite{Correa:2018lyl}.}
\begin{equation}
    \label{bg:J.def}
    J = \int\limits_0^{\psi_0} \frac{C\sin\psi_0\sin^2\psi \, \rmd\psi}{\sqrt{(\sin^2\psi_0-\sin^2\psi)(1+C^2\sin^2\psi_0\sin^2\psi)}}~.
\end{equation}
Evidently, $R_+>R_-$ for $C>0$. Replacing $C$ by $-C$ exchanges $R_+$ and $R_-$. The result \eqref{bg:R.frac} can be generalized to all conformally equivalent configurations by expressing the left hand side in terms of the invariant combination \eqref{conf.invar}. Using a combination that is independent of the sign of $C$, one has  
\begin{equation}
\label{bg:R.frac.gen}
	\frac12 \left( \frac{R_+}{R_-} + \frac{R_-}{R_+} \right) = \frac1{\alpha}~,
\end{equation}
so that \eqref{bg:R.frac} yields
\begin{equation}
\label{bg:alpha}
	\frac{1}{\alpha} = \cosh (2J)~.
\end{equation}

The displacement $\gamma$ along $S^5$ is obtained after substituting \eqref{bg:rho.psi.def}, \eqref{bg:K.sub} and \eqref{bg:rho.eq2} into \eqref{bg:K.def} and solving it for $\theta'$, 
\begin{equation}
\label{bg:dth}
	\theta' = \frac{|\psi'| K\sin\psi_0}{\sqrt{(\sin^2\psi_0-\sin^2\psi)(1+C^2\sin^2\psi_0\sin^2\psi)}}~.
\end{equation}
Integrating along the entire string gives 
\begin{equation}
\label{bg:gamma}
	\gamma = \int\limits_0^{\psi_0} \frac{2 K\sin\psi_0\, \rmd \psi}{\sqrt{(\sin^2\psi_0-\sin^2\psi)(1+C^2\sin^2\psi_0\sin^2\psi)}}\equiv K \hat{\gamma}  ~.
\end{equation}
We shall return later to the integrals in \eqref{bg:alpha} and \eqref{bg:gamma}.


\subsection{Geometry}

Let us schematically denote the AdS$_5\times S^5$ coordinates by  
$$ X^\mu = \left( z,r,\phi,\vec{x}; \theta, \vec{\varphi}\right)~, $$
where   $\vec{x}=(x_1,x_2)$ are two Euclidean coordinates, and  $\varphi$'s are  coordinates on $S^4$. The semicolon separates the AdS$_5$ from the $S^5$ part.
The tangent vectors on the background string world sheet are 
\begin{align}
\label{bg:tan.t}
	X^\mu_\tau &= \left( z',r',0,\zerovec;\theta',\zerovec\right)~,\\
\label{bg:tan.s}
	X^\mu_\sigma &= \left( 0,0,1,\zerovec;0,\zerovec\right)~.
\end{align}
Using the equations of the previous subsection, the induced metric \eqref{bg:induced.metric} reduces to
\begin{equation}
\label{bg:bg.metric}
	g_{\alpha\beta} = L^2 \cot^2 \psi 
		\begin{pmatrix}
			\frac{\theta'^2}{K^2} & 0 \\ 0& 1 
	\end{pmatrix}~.
\end{equation} 
In the light of this result, we find it henceforth useful to adopt the gauge\footnote{This is fine in the limit $K\to 0$, as can be verified from \eqref{bg:K.def}.}
\begin{equation}
\label{bg:choice.tau}
	\theta = K\tau~, \qquad \tau\in (0,2\tau_0)~,
\end{equation}
so that the induced metric is conformal to the Euclidean metric on the cylinder. For massless scalars on the world-sheet, which are conformally invariant, the only possible parameter would be the height of the cylinder, $2\tau_0=\hat{\gamma}$, where $\hat{\gamma}$ was defined in \eqref{bg:gamma}. This will have implications for some of the correlators in subsequent sections. 

For completeness, in the gauge \eqref{bg:choice.tau}, \eqref{bg:dth} and \eqref{bg:rho.eq2} are simply 
\begin{equation}
\label{bg:diff.new}
	\theta'=K\,, \qquad \rho' = C\rho \sin^2 \psi\, ,\qquad 
	\psi' = \pm \sqrt{\cos^2 \psi - K^2 \sin^2 \psi - C^2 \sin^4 \psi}~.
\end{equation}
As discussed before, the $+$ and $-$ signs apply for $\tau\in (0,\tau_0)$ and $\tau\in (\tau_0,2\tau_0)$, respectively. 

One can proceed to calculate the other geometric quantities characterizing the embedding of the background world-sheet. We follow the prescriptions and notation summarized in appendix~\ref{embed}.  
First, one needs to specify an orthonormal set of normal vectors. We take
\begin{align}
\label{bg:N1}
	N_2^\mu &= \frac{z}{L\sqrt{z'^2+r'^2}} \left( -r',z',0,\zerovec;0,\zerovec\right)~,\\
\label{bg:N2}
	N_3^\mu &= \frac{1}{rL\sqrt{z'^2+r'^2}} \left( Kz^2z',Kz^2r',0,\zerovec;K^2z^2-r^2,\zerovec\right)~,\\
\label{bg:N3}
	N_i^\mu &= \frac{z}{L} \left(0,0,0,\vec{n}_i;0,\zerovec\right)~,~~~~\qquad~~~  i=4,5 \\
\label{bg:N5}
    N_i^\mu &= \frac{1}{L\sin\theta} \left(0,0,0,\zerovec;0,\vec{e}_i\right)~,\qquad i=6,7,8,9
\end{align}
where $\vec{n}_i$ and $\vec{e}_i$ denote orthonormal bases on the (Lorentzian) 2-plane and on a unit $S^4$, respectively. 
For convenience, we shall use the indices $i=(2,3,\ldots,9)$ for the normal vectors and reserve $(\flati{0,1})=(\flati{\tau,\sigma})$ for the flat world-sheet indices. 

The second fundamental forms, $H^i{}_{\alpha\beta}$, are determined by the equation of Gauss. The result is 
\begin{align}
\label{bg:H2}
	H^2{}_{\alpha\beta} &= \frac{LC}{\sqrt{1-K^2\tan^2\psi}} 
	\begin{pmatrix} 1 & 0 \\ 0 & -1 \end{pmatrix}~,\\
\label{bg:H3}
	H^3{}_{\alpha\beta} &= \frac{KL\psi'}{\sin\psi\cos\psi\sqrt{1-K^2\tan^2\psi}} 
	\begin{pmatrix} -1 & 0 \\ 0 & 1 \end{pmatrix}~,
\end{align}
and all others vanish. They are, of course, traceless. 

The equation of Weingarten determines the connections in the normal bundle, $A^i{}_{j\alpha}$. It turns out that the only non-vanishing connection is 
\begin{equation}
\label{bg:Aconn}
	A^2{}_{3\tau} = - A^3{}_{2\tau} = \frac{KC\tan^2 \psi}{1-K^2\tan^2\psi}~.
\end{equation} 


\subsection{Renormalized on-shell action}
\label{bg:sec.on.shell.action}
The on-shell action must be regularized and renormalized to obtain a sensible value. In \cite{Correa:2018lyl} this was done by a simple subtraction of the leading divergent term. Here, we prefer to use a covariant prescription adding to \eqref{bg:NG.action}, at the two boundaries, the counter term 
\begin{equation}
\label{bg:ct.action}
	S_{c.t.} = - \sgn z'\, z\, \frac{\delta S_{NG}}{\delta z'}~,
\end{equation}
with $S_{NG}$ given by \eqref{bg:NG.action}. Using the background relations and the gauge \eqref{bg:choice.tau}, the regularized action  becomes
\begin{equation}
\label{bg:reg.action}
	S_{reg} = \sqrt{\lambda} \left[ \int\limits_{\epsilon}^{2\tau_0-\epsilon} \rmd \tau \cot^2 \psi - \left[ \sgn z'\left( \cot\psi\, \psi' +C \sin^2 \psi \right)\right]_{\epsilon}^{2\tau_0-\epsilon}\right]~.
\end{equation}
In particular, we have $\sgn z' =+1$ at $\tau=0$ and $\sgn z' =-1$ at $\tau=2\tau_0$. Therefore, the second part of the counter term cancels between the two boundaries (it also would vanish, because $\sin\psi\to0$ for $\epsilon\to 0$). Moreover, one can show that 
\begin{equation}
\label{bg:cot.psi}
	\cot^2 \psi = -\partial_\tau \left( \cot \psi\, \psi' \right) -( \psi')^2 -C^2 \sin^2 \psi \cos^2\psi~,
\end{equation}
so that the total derivative just cancels the counter term. The integral of the remainder is finite when the cut-off is removed. Thus, the renormalized on-shell action is
\begin{align}
\notag
	S_{ren} &= - \sqrt{\lambda} \int\limits_{0}^{2\tau_0} \rmd \tau\left[(\psi')^2 +C^2 \sin^2 \psi \cos^2\psi \right]\\
\label{bg:ren.action}
	&= -2 \sqrt{\lambda} \int\limits_0^{\psi_0} \rmd \psi \Bigg[ \sqrt{1-(K^2+1)\sin^2\psi -C^2 \sin^4\psi} \\
\notag &\quad +
	\frac{C^2 \sin^2\psi \cos^2\psi}{\sqrt{1-(K^2+1)\sin^2\psi -C^2 \sin^4\psi}}\Bigg]~.
\end{align}

For completeness of presentation, we wish to repeat the study \cite{Correa:2018lyl} of the on-shell action as a function of the macroscopic parameters $\gamma$ and $J$. In \cite{Correa:2018lyl}, the parameters  
\begin{equation}
\label{bg:s.t.def}
	s = \sin^2 \psi_0~, \qquad t = C^2 \sin^4 \psi_0~,
\end{equation}
were introduced, with the parameter space limited by a triangle, $0\leq t \leq 1-s \leq 1$.
The relation \eqref{bg:K.sub} implies 
\begin{equation}
\label{bg:Kst}
    K^2 = \frac{1-s-t}{s}~.
\end{equation}
After changing the integration variable to
\begin{equation}
\label{bg:x.def}
	x= \frac{\sin^2 \psi}{\sin^2\psi_0}~,
\end{equation}
equations \eqref{bg:gamma} and \eqref{bg:J.def} become complete elliptic integrals with modulus\footnote{In order to avoid confusion about the notation, we omit the modulus as the argument of the complete elliptic integrals. Standard references \cite{Gradshteyn, Byrd} use the modulus $k$ as argument, software packages like Mathematica use $m=k^2$; $m$ was used in \cite{Correa:2018lyl}.} 
\begin{equation}
\label{bg:modulus}
	k = \sqrt{\frac{s+t}{1+t}}~.
\end{equation}
Specifically, one finds
\begin{equation}
\label{bg:dtheta.final}
	\gamma = \frac{K}{\sqrt{t}} \int\limits_0^1 \frac{\rmd x}{\sqrt{x(1-x)(\frac1{s}-x)(x+\frac1{t})}} 
	= 2\sqrt{\frac{1-s-t}{1+t}} \mathbf{K}~,
\end{equation}
and 
\begin{equation}
\label{bg:alpha.final}
	J = \int\limits_0^1 \frac{x\, \rmd x}{\sqrt{x(1-x)(\frac1{s}-x)(x+\frac1{t})}} 
	= \sqrt{\frac{t}{s(1+t)}} \left[ \mathbf{K} -(1-s) \mathbf{\Pi}(s) \right]~.
\end{equation}

Similarly, the renormalized on-shell action \eqref{bg:ren.action} becomes
\begin{align}
\notag
	S_{ren} &= -\sqrt{\lambda} \sqrt{t} \int\limits_0^1 \rmd x\left[ \sqrt{\frac{(1-x)(\frac1{t}+x)}{x(\frac1{s}-x)}} + \sqrt{\frac{x(\frac1{s}-x)}{(1-x)(\frac1{t}+x)}} \right] \\
\label{bg:S.ren}
	&= -2\sqrt{\lambda} \sqrt{\frac{1+t}{s}} \left[ \mathbf{E} - (1-k^2)\mathbf{K} \right]~.
\end{align}

\begin{figure}[!th]
\centering
\includegraphics[width=\textwidth]{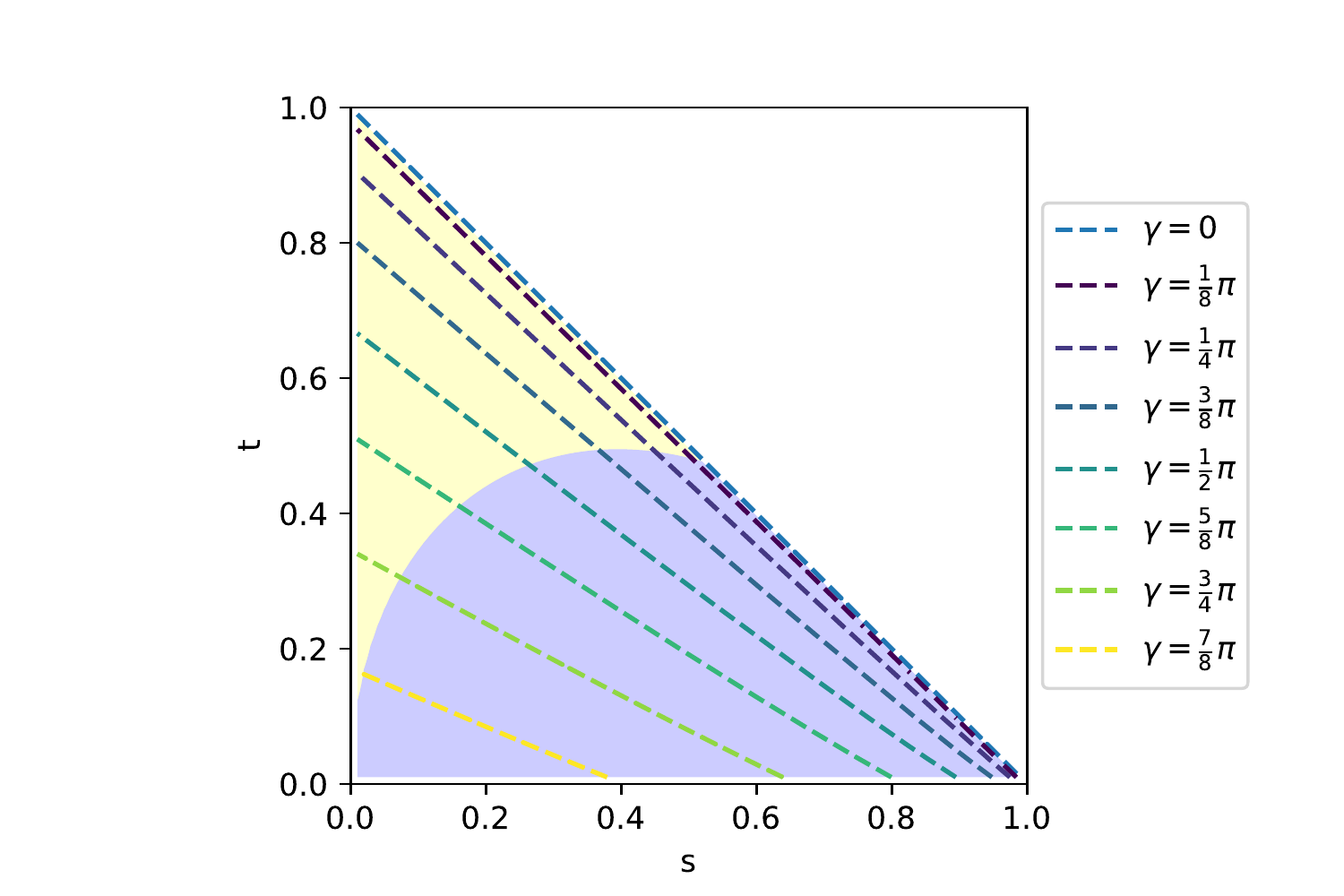}
 \caption{Phase diagram of the Wilson loop correlators in the $s$-$t$-plane. The parameter space for connected solutions is the triangle $s,t\geq0$, $s+t\leq 1$. Connected solutions within the yellow shaded area have $S_{ren}<-2\sqrt{\lambda}$ and are stable. In the blue shaded area, the disconnected solution with $S_{ren}=-2\sqrt{\lambda}$ is the preferred configuration. The disconnected solution is represented by the point $(s,t)=(1,0)$. We remark that the curves with constant $\gamma$ are not straight lines.
 \label{Fig:Phases2}}
\end{figure}

These results agree completely with those of \cite{Correa:2018lyl}, despite the fact that here we have used a different regularization of the action. The resulting phase diagram is illustrated in Figure~\ref{Fig:Phases2} (see also \cite{Correa:2018lyl}). We add some comments on it here. First, the upper limit of the parameter domain, $s+t=1$, corresponds to $K=\gamma=0$, which is the original Wilson loop correlator \cite{Zarembo:1999bu, Olesen:2000ji}. On both axes we have $J=0$, which corresponds to two coincident Wilson loops with different   coupling to the scalars. $s=0$ is the stable branch, whereas $t=0$, which implies $C=0$, is the unstable branch. The point $s=t=0$ is the only point with $\gamma=\pi$, but it is a singular point, because $S_{ren}(0,t)=-\infty$, whereas $\lim\limits_{s\to 0}S_{ren}(s,0) =0$. For $(s,t)=(1,0)$, the connected solution degenerates into two disconnected world-sheets, which barely touch each other, with $S_{ren}=-2\sqrt{\lambda}$. There is a first-order Gross-Ooguri  \cite{Gross:1998gk} phase transition between the connected and the disconnected configurations. This phase transition resonates with some of the arguments involved in the recent replica analysis of Hawking radiation (see \cite{Almheiri:2020cfm} for a review). 

\section{String world-sheet fluctuations}\label{Sec:StringFluctuations}
\subsection{Scalars}
\label{fe:scalars} 

With the scalars $\chi^i$ parameterizing the fluctuations in the normal directions, expanding the Nambu-Goto action to quadratic order around the background gives the action
\begin{equation}
\label{fe:NG.action}
	S_{B} = \frac1{4\pi\alpha'}\int\rmd^2 \sigma \sqrt{g} \left[ (\hat{\nabla}_\alpha \chi_i)(\hat{\nabla}^\alpha \chi^i) - \left( H_{i\alpha\beta} H_j{}^{\alpha\beta} + R_{mpnq} x^{\alpha m} x_{\alpha}^n N_i^p N_j^q \right) \chi^i \chi^j \right]~.
\end{equation}

Hence, the field equations read
\begin{equation}
\label{fe:eom}
	\left[ \delta^i_j \hat{\nabla}^\alpha \hat{\nabla}_\alpha + H^i{}_{\alpha\beta} H_j{}^{\alpha\beta} - M^i{}_j \right] \chi^j =0~,
\end{equation}
with 
\begin{equation}
\label{fe:M.def}
	M^i{}_j = -R_{\lambda\mu\nu\rho} g^{\alpha\beta} X_\alpha^\lambda X_\beta^\nu N^{i\mu}N_j{}^\rho~. 
\end{equation}
Note that $\hat{\nabla}_\alpha$ is the generalized covariant derivative \eqref{embed:nabla.ex}.
Using the background relations, we have explicitly 
\begin{align}
\label{fe:M11}
	M^2{}_2 &= M^i{}_i=\frac2{L^2} - \frac{K^2}{L^2} \tan^2\psi &&(i=4,5)~,\\
\label{fe:M22}
	M^3{}_3 &= -M^i{}_i= \frac{K^2}{L^2} \tan^2\psi &&(i=6,7,8,9)~,
\end{align}
and all the off-diagonal elements vanish.

The field equations \eqref{fe:eom} can be written down straightforwardly. After some manipulations, they read explicitly
\begin{multline}
\label{fe:eq1}
	\left[\frac{\partial^2}{\partial \tau^2} + \frac{\partial^2}{\partial \phi^2} +\frac4{\sin^2\psi} 
	- \frac{\tan^2\psi}{\sqrt{1-K^2\tan^2\psi}} \left( \frac{\sqrt{1-K^2\tan^2\psi}}{\tan^2\psi} \right)'' \right] \chi^2 
		= \\ -\frac{2KC\tan^2\psi}{\sqrt{1-K^2\tan^2\psi}} 
		\frac{\partial}{\partial \tau}\left(\frac{\chi^3}{\sqrt{1-K^2\tan^2\psi}}\right)~,
\end{multline}
\begin{multline}
\label{fe:eq2}
	\left[\frac{\partial^2}{\partial \tau^2} + \frac{\partial^2}{\partial \phi^2} 
	- \frac1{\sqrt{1-K^2\tan^2\psi}} \left( \sqrt{1-K^2\tan^2\psi}\right)'' \right] \chi^3 
		= \\ \frac{2KC}{\sqrt{1-K^2\tan^2\psi}} 
		\frac{\partial}{\partial \tau} \left(\frac{\chi^2\tan^2\psi}{\sqrt{1-K^2\tan^2\psi}}\right)~,
\end{multline}
and 
\begin{align}
\label{fe:eq3}
	\left(\frac{\partial^2}{\partial \tau^2} + \frac{\partial^2}{\partial \phi^2} -2 \cot^2 \psi + K^2 \right) \chi^i &=0 &&(i=4,5)~,\\
\label{fe:eq5}
	\left(\frac{\partial^2}{\partial \tau^2} + \frac{\partial^2}{\partial \phi^2} + K^2 \right) \chi^i &=0 &&(i=6,7,8,9)~.  
\end{align}
The above equations are consistent with the background world-sheet symmetry $\tau \to 2\tau_0-\tau$, because $K$ and $C$ change sign under this symmetry. This implies that $N_2^\mu$ flips its direction while $N_3^\mu$ remains invariant. As a consequence, we also have $\chi^2\to -\chi^2$.  

Let us make a few remarks that somewhat clarify the structure of the fluctuations in various limits and will provide a blueprint for field theory expectations.

\begin{itemize}

\item An important point of reference are the fluctuations of the half-BPS Wilson-Maldacena loop. In that case the fluctuations have a natural 5+3 split which is interpreted as corresponding to the protected operators: $\Phi^{a=1,2,3,4,5}$ with $\Delta=1$ and $F_{ti}+iD_i\Phi^6$, $i=1,2,3$ with $\Delta=2$. We now recognize that the last four fluctuations presented in equation \eqref{fe:eq5} are the string theoretic scalar modes dual to the four operators constructed from $\Phi^{a=1,2,3,4}$.  Analogously, the fluctuations in equation \eqref{fe:eq3}, correspond to two operators with $\Delta=2$. The simplicity of these two sets of equations of motion suggests that the field theory treatment might be manageable.

\item  In the special case $K =0$ (with arbitrary $C$), the bosonic modes organize themselves according to a 5+2+1 split. First, the mode $\chi^3$ joins the four mode $\chi^{i=6,7,8,9}$ forming a quintet of massless scalars. This is intuitively clear, because with $K=0$ implies no displacement of the classical world-sheet along the $S^5$.  For the mode $\chi^2$ one can check that  
\begin{eqnarray}
    \frac{4}{\sin^2\psi}-\tan^2\psi(\cot^2\psi)'' &\stackrel{K\to 0 }{\longrightarrow}   & -2\cot^2\psi  +2C^2\tan^2 \psi.
\end{eqnarray}
The last term above,  $2C^2\tan^2\psi$, is sub-leading near the boundary and we see how mode $\chi^2$ almost pairs with the modes $\chi^{i=4,5}$.

\item Let us now consider the $C\to 0$ limit. The fluctuation $\chi^3$ does not obey the same equations as $\chi^{i=6,7,8,9}$ as there is motion on $S^5$.  However, we do expect, given that $C=0$ corresponds to a classical world-sheet that extends only along $S^5$,
the fluctuation $\chi^2$ to satisfy the same equation as fluctuations $\chi^{i=4,5}$. Indeed, 
\begin{eqnarray}
\frac{4}{\sin^4\psi}-\frac{\tan^2\psi}{\sqrt{1-K^2\tan^2\psi}}\left( \frac{\sqrt{1-K^2\tan^2\psi}}{\tan^2\psi} \right)'' & \stackrel{C\to 0}{\longrightarrow} & -2\cot^2\psi +K^2.
\end{eqnarray}

\item Equations \eqref{fe:eq1} and  \eqref{fe:eq2} decouple for either  $C=0$ or $K=0$, which geometrically correspond to world-sheets that stay strictly within $S^5$ or AdS$_5$, respectively. In the general configuration, the coupling between $\chi^2$ and $\chi^3$ suggests, on the field theory side, an unexpected mixing between operators which in the half-BPS limit had conformal dimensions $\Delta=1$ and $\Delta=2$. This mixing is generated by the breaking of conformal invariance.

\end{itemize}


\subsection{Fermions}
\label{ferm}

The part of the type IIB superstring action, which is quadratic in the fermions, is given by \cite{Metsaev:2001bj,Drukker:2000ep}
\begin{equation}
\label{ferm:action}
	S_F = \frac{1}{2\pi\alpha'} \int \rmd^2 \xi\, \sqrt{g}\, \bar{\Theta} \left( g^{\alpha\beta} -i\epsilon^{\alpha\beta} \sigma_3 \right) \Gamma_\alpha \mathcal{D}_\beta \Theta~.
\end{equation}
This is the action for a Euclidean world sheet, which is appropriate in our case. Notice that this is not in contradiction with the fact that the bulk is Lorentzian AdS$_5\times S^5$. 
We use the double spinor notation \cite{Martucci:2005rb}, so that $\Theta$ is a 64-component spinor consisting of two positive-chirality 10-d Majorana-Weyl spinors, with the explicit Pauli matrices acting on the spinor doublet. In contrast to \cite{Drukker:2000ep}, our $\epsilon^{\alpha\beta}$ is the epsilon tensor, not a density. Moreover, in \eqref{ferm:action}, the generalized covariant spinor derivative $\mathcal{D}_\alpha$ is  \cite{Martucci:2005rb}
\begin{equation}
\label{ferm:cov.deriv}
	\mathcal{D}_\alpha  = \hat{D}_\alpha  +\frac1{16} \slashed{F} \Gamma_\alpha (i\sigma_2)~,
\end{equation}
with
\begin{equation}
\label{ferm:slashF}
	\slashed{F} = \frac1{5!} F_{pqrst} \Gamma^{pqrst}~,
\end{equation}
and $\hat{D}_\alpha$ is the pull-back of the bulk covariant derivative on the world sheet given by \eqref{embed:D}. 

Our first aim is to rewrite the action \eqref{ferm:action} in terms of eight genuine Euclidean 2-d spinors, which would be the super partners of the eight scalars, if the background were supersymmetric.  
A number of steps are necessary to achieve this aim, first of all $\kappa$ symmetry gauge fixing, which reduces the $16+16$ components of $\Theta$ (Although the double spinor has 64 components, half of them vanish by the chirality condition.) to $16=8\times 2$. 

Because, in two dimensions, $\epsilon^{\alpha\beta} \Gamma_\alpha = \Gamma^{\flati{01}} \Gamma^\beta$,  the action in \eqref{ferm:action} simplifies to 
\begin{equation}
\label{ferm:action2}
	S_F = \frac{1}{2\pi\alpha'} \int \rmd^2 \xi\, \sqrt{g}\, \bar{\Theta} \left( 1 - i \sigma_3 \Gamma^{\flati{01}} \right) \Gamma^\alpha \mathcal{D}_\alpha \Theta~, 
\end{equation}
which enjoys the $\kappa$ symmetry 
\begin{equation}
\label{ferm:kappa.sym}
	\delta \Theta = \frac12 \left( 1 -i \sigma_3 \Gamma^{\flati{01}} \right)\kappa~.
\end{equation}
This allows to fix
\begin{equation}
\label{ferm:kappa.fix}
	\Theta^1=\Theta^2 \equiv \Theta~.
\end{equation}
Thus, henceforth, $\Theta$ is a single positive-chirality 10-d spinor, and the action \eqref{ferm:action2} reduces to
\begin{equation}
\label{ferm:action3}
	S_F = \frac{1}{\pi\alpha'} \int \rmd^2 \xi\, \sqrt{g}\, \bar{\Theta}  
	\left( \Gamma^\alpha \hat{D}_\alpha 
	- \frac{i}{16} \Gamma^{\flati{01}} \Gamma^\alpha \slashed{F} \Gamma_\alpha \right) \Theta~.
\end{equation}
Let us consider the terms in the parentheses separately. First, using the fact that the second fundamental forms $H^i_{\alpha\beta}$ are traceless on the background world sheet, the  corresponding term in $\hat{D}_\alpha$ \eqref{embed:D} vanishes, so that 
\begin{equation}
\label{ferm:D}
	\Gamma^\alpha \hat{D}_\alpha = \Gamma^\alpha D_\alpha +\frac14 A_{ij\alpha} \Gamma^\alpha \Gamma^{ij}~,
\end{equation}
where $D_\alpha$ is the standard 2-d (covariant) spinor derivative and the normal bundle connection is given by \eqref{bg:Aconn}. 

To obtain the contraction $\Gamma^\alpha \slashed{F} \Gamma_\alpha$ in \eqref{ferm:action3}, we first decompose $\slashed{F}$ into tangential and normal components using the completeness relation \eqref{embed:ortho} by writing
$$  \slashed{F} = \frac1{5!} \left(F_{i_1\cdots i_5} \Gamma^{i_1\cdots i_5} + 5 F_{\beta i_1\cdots i_4} \Gamma^{\beta i_1\cdots i_4} 
+10 F_{\beta_1 \beta_2  i_1 i_2i_3} \Gamma^{\beta_1 \beta_2  i_1 i_2 i_3} \right)~.
$$
The terms with more than two tangential components vanish by antisymmetry. After carrying out the contractions and using $\Gamma^{\alpha\beta}=\epsilon^{\alpha\beta} \Gamma^{\flati{01}}$, we get
\begin{equation}
\label{ferm:F.contract}
	\Gamma^\alpha \slashed{F} \Gamma_\alpha = -\frac2{5!} F_{i_1\cdots i_5} \Gamma^{i_1\cdots i_5} + \frac1{3!} \epsilon^{\alpha\beta} 
	F_{\alpha \beta  i_1 i_2 i_3} \Gamma^{\flati{01}} \Gamma^{i_1 i_2 i_3}~.
\end{equation}

To make progress, we need to substitute five-form field strength \eqref{ferm:F5}.
In the calculation of the Hodge dual, there is a subtlety related to the frame orientation, which is, in turn, related to the chirality matrix. In order to make this explicit, we take the volume form of the bulk to be 
\begin{equation}
	\epsilon = \pm \epsilon_{AdS} \wedge \epsilon_{S^5}~,
\end{equation}  
so that \eqref{ferm:F5} is
\begin{equation}
\label{ferm:F5.expl}
	F_5 = \frac4{L} \left(\mp\epsilon_{AdS} +\epsilon_{S^5} \right)~.
\end{equation}
We can see that the first term on the right hand side of \eqref{ferm:F.contract} can receive a contribution only from the $S^5$ part of $F_5$, because only on $S^5$ there are non-zero components of five normal vectors ($i= 3,6,7,8,9$). Similarly, the second term arises from the AdS part of $F_5$, because it needs two non-zero components of the tangents, while the normals involved are $i=2,4,5$, because the AdS part of $N_3^\mu$ is proportional to the AdS part of the tangent $X_\tau^\mu$. Explicitly, one finds (recall that the normal indices are flat)
\begin{equation}
\label{ferm:Fc}
	\Gamma^\alpha \slashed{F} \Gamma_\alpha = \frac{8\sqrt{1-K^2\tan^2\psi}}{L} 
    \left( \Gamma^{36789} \pm \Gamma^{\flati{01}245} \right)~, 
\end{equation} 
so that 
\begin{equation}
\label{ferm:Fc2}
	\Gamma^{\flati{01}} \Gamma^\alpha \slashed{F} \Gamma_\alpha = \mp \frac{8\sqrt{1-K^2\tan^2\psi}}{L} 
    \Gamma^{245}\left( 1 \mp \Gamma^{\flati{01}23456789} \right)~. 
\end{equation} 
In these calculations, recall that the world sheet is Euclidean, whereas the normal vector $N^\mu_5$ is time-like. This implies $(\Gamma^5)^2= (\Gamma^{\flati{01}})^2 = -1$. 
By convention, the right hand side of \eqref{ferm:Fc2} must project onto positive-chirality spinors, so that $\Theta$ in \eqref{ferm:action3} is projected onto itself. Taking the 10-d chirality matrix to be ($5$ is the bulk time direction)
\begin{equation}
\label{ferm:chir.mat}
	\Gamma_{(10)}= \Gamma^{5\flati{01}2346789} = - \Gamma^{\flati{01}23456789} 
\end{equation}
implies the upper sign in the previous equations. Of course, the opposite choice implying the lower sign is equivalent.

Moreover, it is convenient to write the action in a form that is more appropriate to the Euclidean world sheet and explicitly remove the 16 negative-chirality components of $\Theta$. Recalling that $i=5$ corresponds to the bulk time direction, the conjugate spinor is $\bar{\Theta} = \Theta^\dagger \Gamma^5$. In the action \eqref{ferm:action3}, let us move that $\Gamma^5$ from $\bar\Theta$ into the operator to the right of it and define $\gamma^\alpha = \Gamma^5 \Gamma^\alpha$ and $\gamma^i = \Gamma^5 \Gamma^i$ ($i\neq 5$). These matrices span a Euclidean 9-d Clifford algebra and act freely on the 16-component spinor that is equivalent to the 10-d positive-chirality spinor $\Theta$.\footnote{In particular, the representation is such that $\gamma^{\flati{01}2346789}=1$.} 
Therefore, after substituting \eqref{ferm:D}, together with \eqref{bg:Aconn}, and \eqref{ferm:Fc2} with the upper sign into \eqref{ferm:action3} and transforming the gamma matrices, we obtain
\begin{equation}
\label{ferm:action4}
	S_F = \frac{1}{\pi\alpha'} \int \rmd^2 \xi\, \sqrt{g}\, \Theta^\dagger  
	\left( \gamma^\alpha D_\alpha +\frac{KC \tan^3\psi}{2L(1-K^2\tan^2\psi)} \gamma^{\flati{0}} \gamma^{23} 
	- \frac{i}{L} \sqrt{1-K^2\tan^2\psi} \gamma^{24} \right) \Theta~,
\end{equation}   
where $\Theta$ is now a 16-component spinor.

It remains to decompose $\Theta$ into eight 2-d spinors. This can be achieved by introducing three quantum numbers, $a,b,c=\pm 1$, and taking $\Theta^{abc}$ such that
\begin{equation}
\label{ferm:Theta.abc}
	\gamma^{24} \Theta^{abc} = ia \Theta^{abc}~,\qquad
	\gamma^{67} \Theta^{abc} = ib \Theta^{abc}~,\qquad
	\gamma^{89} \Theta^{abc} = ic \Theta^{abc}~.	
\end{equation} 
The action of $\gamma^{23}$ and $\gamma^{34}$ can be obtained by noting that $[\gamma^{23}, \gamma^{34}] = 2 \gamma^{24}$. A suitable, albeit not unique, choice is
$$ \gamma^{23} \Theta^{abc} = \Theta^{-abc}~,\qquad \gamma^{34} \Theta^{abc} = ia \Theta^{-abc}~.$$
After this decomposition, \eqref{ferm:action4} finally reduces to 
\begin{align}
\label{ferm:action5}
	S_F &= \frac{1}{\pi\alpha'} \int \rmd^2 \xi\, \sqrt{g}\, \Theta^{abc}{}^\dagger  
	\Bigg[ \left(\gamma^\alpha D_\alpha + \frac{a}{L} \sqrt{1-K^2\tan^2\psi} \right) \Theta^{abc} \\
\notag    &\quad
	+ \frac{KC \tan^3\psi}{2L(1-K^2\tan^2\psi)} \gamma^{\flati{0}} \Theta^{-abc} \Bigg]~,
\end{align}  
where spinors can be treated as 2-d spinors, and the sum over $a,b$, and $c$ is implicit.

In what follows, we shall drop the labels $b$ and $c$, because they do not appear as parameters in the field equations. The derivative operator is 
\begin{equation}
\label{ferm:D.expl}
	\gamma^\alpha D_\alpha = \gamma^\alpha \partial_\alpha -\frac{\psi'}{2L\cos^2\psi} \gamma^{\flati{0}}~, 
\end{equation}
so that the field equations are
\begin{equation}
\label{ferm:eom}
	\left( \partial_\tau + \gamma^{\flati{01}} \partial_\sigma - \frac{\psi'}{2\sin\psi\cos\psi} + \frac{a\sqrt{1-K^2\tan^2\psi}}{\tan\psi} \gamma^{\flati{0}} \right) \Theta^a +  
	\frac{KC \tan^2\psi}{2(1-K^2\tan^2\psi)} \Theta^{-a} = 0~. 
\end{equation} 

Our intuition of the fermionic fluctuations is not as clear as the one for bosonic fluctuations. Notice that the above expression for the fluctuations is independent of the $b$ and $c$ indices and points to four couples of inter-related 2-d fermions. Either in the limit of vanishing $K$ or vanishing $C$, the fermion field equations decouple into $4+4$ equations, somewhat mirroring the situation for the bosonic fields.

\section{Correlators}\label{Sec:Correlators}

In this section we employ the AdS/CFT correspondence to study the strongly coupled limit of correlators of operators dual to the string theoretic fluctuations discussed in the previous section. We start with a concise discussion of the holographic calculation with Dirichlet boundary conditions and how this prescription is modified by the presence of two boundaries. Then we discuss the massless and  massive scalar cases, and finish the section with an estimate for the correlators of heavy operators which can be computed in a geodesic approximation. 

\subsection{Calculating two-point functions}
\label{ads.cft}
Consider the EAdS$_2$ metric
\begin{equation}
    \label{ads:metric}
    \rmd s^2 = \frac{L^2}{\tau^2} \left(\rmd \tau^2 + \rmd \phi^2\right)~.    
\end{equation}
The boundary $C$ is located at $\tau=\epsilon\to0$. A free massive scalar in this geometry satisfies the field equation
\begin{equation}
    \label{ads:eom}
        (\nabla^2 -m^2) \chi =0\quad \Rightarrow \quad 
        \left( \partial_\tau^2 + \partial_\phi^2  - \frac{m^2 L^2}{\tau^2} \right) \chi =0~.
\end{equation}
Following the Frobenius method, the general asymptotic solution can be written as
\begin{empheq}{alignat=7}
    \label{ads:sol.series}
    \chi(\tau,\phi)&\underset{\tau\to0} {\longrightarrow}\tau^{1-\Delta}\left(1+\cdots\right)\hat{\chi}(\phi)+\tau^\Delta\left(1+\cdots\right)\check{\chi}(\phi)\,,
\end{empheq}
where 
$$ \Delta = \frac12 +\sqrt{\frac14 +m^2 L^2} $$
denotes the conformal dimension of the dual operator. The functions $\hat{\chi}(\phi)$ and $\check{\chi}(\phi)$ are called the source and response function, respectively. We are going to focus on two types of fluctuations: $\chi^{i=6,7,8,9}$ with $m^2L^2=0$ ($\Delta=1$), and $\chi^{i=4,5}$ with $m^2L^2=2$ ($\Delta=2$).

Assuming that the scalar is regular in the interior, there always exist (possibly divergent) boundary terms such that the on-shell variation of the action is finite as $\epsilon\to0$ and reads \cite{Andrade:2006pg}
\begin{equation}
    \label{onshellSvariation}
    \delta S=\left(2\Delta-1\right)\int d\phi\,\check{\chi}(\phi)\delta\hat{\chi}(\phi)\,.
\end{equation}
This is appropriate for the Dirichlet problem in which $\hat{\chi}$ is fixed. Then, according to the AdS/CFT correspondence, the semiclassical one-point function (in the presence of sources) of the operator dual to $\chi$ is \cite{Witten:1998qj,Klebanov:1999tb,Bianchi:2001kw,Marolf:2006nd}
\begin{equation}
    \label{ads:1pt}
    \langle \mathcal{O}(\phi) \rangle = (2\Delta-1) \check{\chi}(\phi)~, 
\end{equation}
expression from which the two-point function is found by differentiation, namely,
\begin{equation}
    \label{ads:2pt}
    \langle \mathcal{O}(\phi_1)\mathcal{O}(\phi_2) \rangle = (2\Delta-1) \frac{\delta\check{\chi}(\phi_1)}{\delta \hat{\chi}(\phi_2)}~. 
\end{equation}
It is important to emphasize that, in the case of a single boundary, regularity of the fields in the interior $\tau\to\infty$ is what determines the vev $\check{\chi}(\phi)$ as a function of the source $\hat{\chi}(\phi)$.

In our case the metric \eqref{bg:bg.metric} approaches the EAdS$_2$ geomtery for $\tau\to0$ as well as for $\tau\to2\tau_0$, and the expansion \eqref{ads:sol.series} applies on both ends, that is,
\begin{empheq}{alignat=7}
    \label{2bdry:sol.seriesL}
    \chi(\tau,\phi)&\underset{\tau\to0}{\longrightarrow}\tau^{1-\Delta}\left(1+\cdots\right)\hat{\chi}^L(\phi)+\tau^\Delta\left(1+\cdots\right)\check{\chi}^L(\phi)\,,
\end{empheq}
and
\begin{empheq}{alignat=7}
    \label{2bdry:sol.seriesR} \chi(\tau,\phi)&\underset{\tau\to2\tau_0}{\longrightarrow}(2\tau_0-\tau)^{1-\Delta}\left(1+\cdots\right)\hat{\chi}^R(\phi)+(2\tau_0-\tau)^\Delta\left(1+\cdots\right)\check{\chi}^R(\phi)\,.
\end{empheq}
A similar analysis as above shows that the variation of the on-shell action reads
\begin{equation}
    \label{onshellSvariation}
    \delta S=\left(2\Delta-1\right)\oint d\phi\,\check{\chi}^L(\phi)\delta\hat{\chi}^L(\phi)+\left(2\Delta-1\right)\oint d\phi\,\check{\chi}^R(\phi)\delta\hat{\chi}^R(\phi)\,.
\end{equation}
This time, however, no regularity condition is required for $\tau\to\infty$ because the coordinate never approaches that value. Instead, knowledge of the two independent sources $\hat{\chi}^{L,R}(\phi)$ is sufficient to yield a regular solution and uniquely determine the response functions $\check{\chi}^{L,R}(\phi)$. In this setup, the AdS/CFT correspondence describes two interacting CFT's living on each boundary with $1$-point functions
\begin{empheq}{alignat=7}
    \label{2bdry:1pt}
    \langle \mathcal{O}^L(\phi) \rangle &=(2\Delta-1) \check{\chi}^L(\phi)\,,
    &\qquad
    \langle \mathcal{O}^R(\phi) \rangle &=(2\Delta-1) \check{\chi}^R(\phi)\,.
\end{empheq}
There are then two types of $2$-point functions we can compute, namely,
\begin{empheq}{alignat=7}
    \label{2bdry:2pt}
    \langle \mathcal{O}^{L,R}(\phi_1)\mathcal{O}^{L,R}(\phi_2) \rangle &= (2\Delta-1) \frac{\delta\check{\chi}^{L,R}(\phi_1)}{\delta\hat{\chi}^{L,R}(\phi_2)}\,,
    \\
    \label{2bdry:2ptmixed}
    \langle \mathcal{O}^{L,R}(\phi_1)\mathcal{O}^{R,L}(\phi_2) \rangle &= (2\Delta-1) \frac{\delta\check{\chi}^{L,R}(\phi_1)}{\delta\hat{\chi}^{R,L}(\phi_2)}\,,
\end{empheq}
depending on whether the insertions lie on the same boundary or not. Finally, since $\phi$ has periodicity $2\pi$, we can Fourier-expand the fields,
\begin{empheq}{alignat=7}
    \label{ads:modes}
    \hat{\chi}^{L,R}(\phi)&=\sum\limits_{n=-\infty}^\infty\e{in\phi}\hat{\chi}_n^{L,R}\,,
    &\qquad
    \check{\chi}^{L,R}(\phi)&=\sum\limits_{n=-\infty}^\infty \e{in\phi}\check{\chi}_n^{L,R}.
\end{empheq}
Modes do not mix for free fields so \eqref{2bdry:2pt}-\eqref{2bdry:2ptmixed} reduce to 
\begin{equation}
    \label{ads:2pt.modes}
    \langle \mathcal{O}^{L,R}(\phi_1)\mathcal{O}^{R}(\phi_2) \rangle = (2\Delta-1)  \sum\limits_{n=-\infty}^\infty \e{in(\phi_1-\phi_2)} \frac{\partial\check{\chi}_n^{L,R}}{\partial\hat{\chi}_n^{R}}~. 
\end{equation}

\subsection{Massless scalar}
For general $K$, none of the scalars discussed in subsection~\ref{fe:scalars} are truly massless. However, in the special case $K=0$ all the fields associated with fluctuations on $S^5$ (the four scalars in \eqref{fe:eq5} and $\chi^3$) become massless. In this subsection we shall consider this special case; for massless fields one can provide a full analytic expression for the correlator while illustrating the method for the general case. 

Because the induced metric \eqref{bg:bg.metric} is conformally flat, the massless field equation is simply 
\begin{equation}
    \left( \frac{\partial^2}{\partial\tau^2} + \frac{\partial^2}{\partial\phi^2} \right) \chi(\tau,\sigma) =0~,
\end{equation}
the general solution of which reads 
\begin{equation}
\label{fe:massless.sol}
    \chi(\tau,\phi) = \sum\limits_{n=-\infty}^{\infty}\left[A_n\cosh\left(n\tau\right)+B_n\sinh\left( n\tau\right)\right]\e{in\phi}~,
\end{equation}
where $A_n$ and $B_n$ are arbitrary coefficients. Expanding close to the boundaries we can read the Fourier coefficients of the source and response functions ($\Delta=1$ for a massless scalar),
\begin{empheq}{alignat=7}
    \hat{\chi}_n^L&=A_n\,,
    &\qquad
    \hat{\chi}^R&=A_n\cosh(2n\tau_0)+B_n\sinh(2n\tau_0)\,,
    \\
    \check{\chi}_n^L&=nB_n\,,
    &\qquad
    \check{\chi}_n^R&=-n\left(A_n\sinh(2n\tau_0)+B_n\cosh(2n\tau_0)\right)\,,
\end{empheq}
from where we solve
\begin{empheq}{alignat=7}
    \check{\chi}_n^{L,R}&=\frac{n}{\sinh(2n\tau_0)}\left(-\cosh(2n\tau_0)\hat{\chi}_n^{L,R}+\hat{\chi}_n^{R,L}\right)\,.
\end{empheq}
Formula \eqref{ads:2pt.modes} then yields the correlation functions
\begin{empheq}{alignat=7}
\notag
	\langle\mathcal{O}^{L,R}(\phi)\mathcal{O}^{L,R}(0)\rangle&=-\sum_{n=-\infty}^{\infty}n\coth(2\tau_0\, n)\e{in\phi}
	\\
\label{fe:OLL.massless}
    &=-\frac{1}{2\tau_0}-2\sum_{n=1}^{\infty}n\coth(2\tau_0\, n)\cos\left(n\phi\right)\,,
	\\
 \notag
	\langle\mathcal{O}^{L,R}(\phi)\mathcal{O}^{R,L}(0)\rangle&=\sum_{n=-\infty}^{\infty}\frac{n}{\sinh(2\tau_0\, n)}\e{in\phi}
	\\
\label{fe:OLR.massless}
    &=\frac{1}{2\tau_0}+2\sum_{n=1}^{\infty}\frac{n}{\sinh(2\tau_0\, n )}\cos\left(n\phi\right)\,.
\end{empheq}
To compute these sums we recall the following properties of the $\theta$ functions:
\begin{empheq}{alignat=7}
	\frac{\partial}{\partial z}\ln\theta_1(z,q)&=\cot z+4\sum_{n=1}^{\infty}\frac{q^n}{q^{-n}-q^n}\sin(2nz)\,,
	\\
	\frac{\partial}{\partial z}\ln\theta_4(z,q)&=4\sum_{n=1}^{\infty}\frac{1}{q^{-n}-q^n}\sin(2nz)\,.
\end{empheq}
It  follows that
\begin{empheq}{alignat=7}
\label{fe:theta1}
	\frac{\partial^2}{\partial z^2}\ln\theta_1\left(z,q\right)&=-\frac{1}{\sin^2 z} + 8 \sum_{n=1}^{\infty}\frac{nq^n}{q^{-n}-q^n}\cos(2nz)\,,
	\\
	\frac{\partial^2}{\partial z^2}\ln\theta_4\left(z,q\right)&=8\sum_{n=1}^{\infty}\frac{n}{q^{-n}-q^n}\cos(2nz)\,.
\end{empheq}
Using the representation
\begin{empheq}{alignat=7}
	-\frac{1}{\sin^2 z}&=4\sum_{n=1}^{\infty}n\cos(2nz)\,,
\end{empheq}
\eqref{fe:theta1} becomes 
\begin{empheq}{alignat=7}
	\frac{\partial^2}{\partial z^2}\ln\theta_1\left(z,q\right)&=4\sum_{n=1}^{\infty}\frac{q^{-n}+q^n}{q^{-n}-q^n}n\cos(2nz)\,.
\end{empheq}
Thus, we can re-express the two-point functions \eqref{fe:OLL.massless} and \eqref{fe:OLR.massless} as 
\begin{empheq}{alignat=7}
\label{fe:OLL.massless.fin}
	\langle\mathcal{O}^{L,R}(\phi)\mathcal{O}^{L,R}(0)\rangle&=-\frac{1}{2\tau_0}-2\frac{\partial^2}{\partial\phi^2}\ln\theta_1\left(\frac{\phi}{2},\e{-2\tau_0 }\right)\,,
	\\
\label{fe:OLR.massless.fin}
    \langle\mathcal{O}^{L,R}(\phi)\mathcal{O}^{R,L}(0)\rangle&=\frac{1}{2\tau_0}+2\frac{\partial^2}{\partial\phi^2}\ln\theta_4\left(\frac{\phi}{2},\e{-2\tau_0}\right)\,.
\end{empheq}

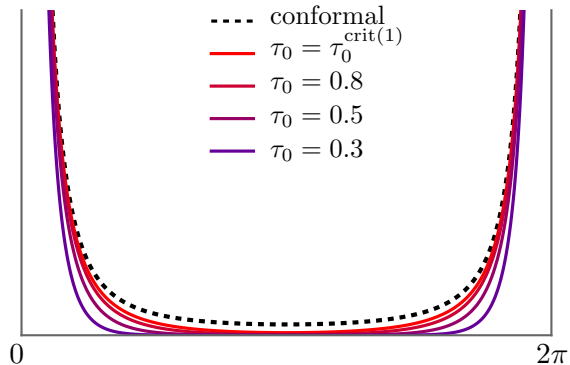
\begin{figure}[!th]
\centering
\begin{tikzpicture}
\node at (-5.45,-0.85) {\includegraphics[width=7cm]{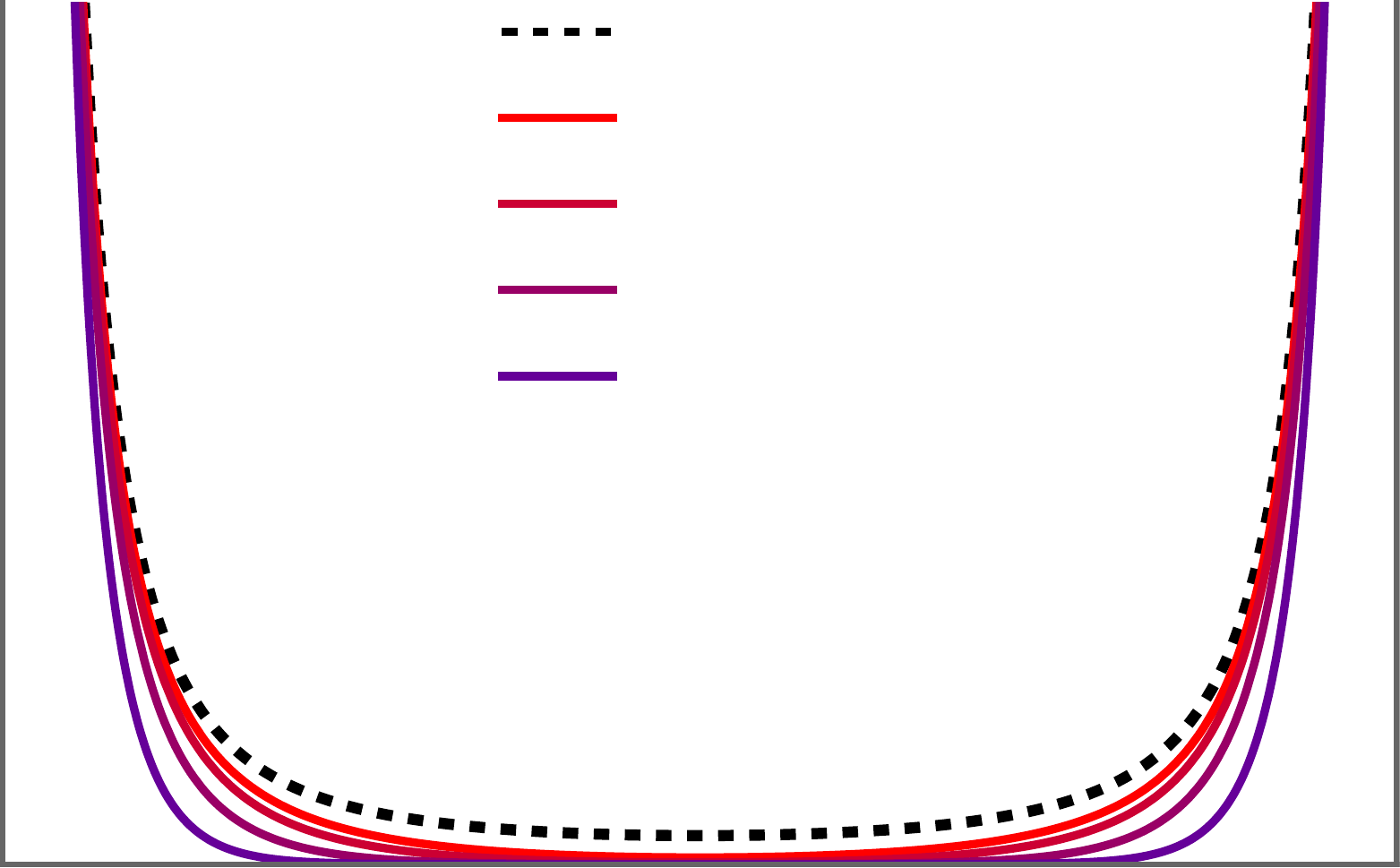}};
\node at (-1.95,-3.25) {$2\pi$};
\node at (-9,-3.25) {$0$};
\node[right] at (-5.8,1.25) {\small conformal};
\node[right] at (-5.8,0.85) {\small $\tau_0 = \tau_0^{\rm crit(1)}$};
\node[right] at (-5.8,0.35) {\small $\tau_0 = 0.8$};
\node[right] at (-5.8,-0.1) {\small $\tau_0 = 0.5$};
\node[right] at (-5.8,-0.55) {\small $\tau_0 = 0.3$};
\end{tikzpicture}
\caption{$\langle\mathcal{O}^{L}(\phi)\mathcal{O}^{L}(0)\rangle$ correlator as a function of $\phi\in (0,2 \pi)$ for values of $\tau_0$ below and including the critical one $\tau_0^{{\rm crit}(1)}=1.218$}
\label{plotLL}
\end{figure}

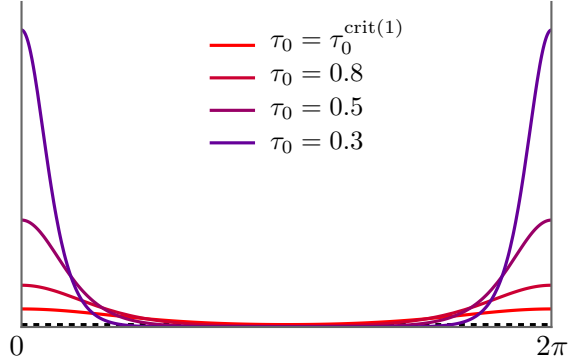
\begin{figure}[!ht]
\centering
\begin{tikzpicture}
\node at (-5.45,-0.85) {\includegraphics[width=7cm]{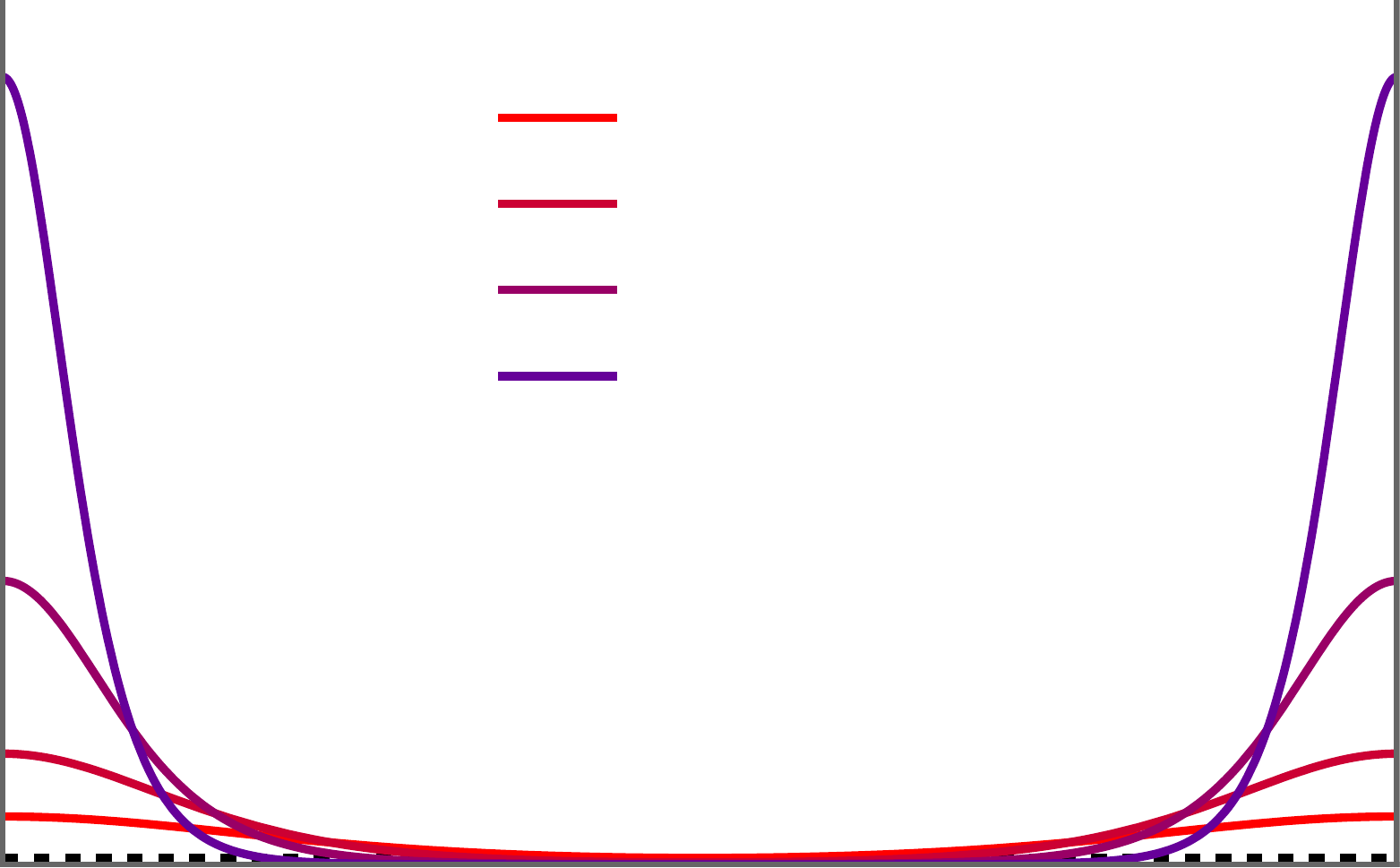}};
\node at (-1.95,-3.25) {$2\pi$};
\node at (-9,-3.25) {$0$};
\node[right] at (-5.8,0.85) {\small $\tau_0 = \tau_0^{\rm crit(1)}$};
\node[right] at (-5.8,0.35) {\small $\tau_0 = 0.8$};
\node[right] at (-5.8,-0.1) {\small $\tau_0 = 0.5$};
\node[right] at (-5.8,-0.55) {\small $\tau_0 = 0.3$};
\end{tikzpicture}
\caption{$\langle\mathcal{O}^{L}(\phi)\mathcal{O}^{R}(0)\rangle$ correlator as a function of $\phi\in (0,2 \pi)$  for values of $\tau_0$ below and including the critical one $\tau_0^{{\rm crit}(1)}=1.218$ }
\label{plotLR}
\end{figure}

Let us finish this subsection by addressing  some of the key properties of the two-point functions. A convenient way to visualize the information is provided in Figures \ref{plotLL} and \ref{plotLR}, where we have plotted the correlators for different values of $\tau_0$.    The chosen values include points near the Gross-Ooguri phase transition. Recall that the correlators were obtained in the $K=0$ case which, according to Eq. \eqref{bg:Kst}, implies $1=t+s$ corresponding to the dotted blue line in the phase diagram Figure~\ref{Fig:Phases2}. To determine the precise values of $\tau_0$ we translate to the parameters $(s,t)$. The above situation corresponds to $K=0$, then $C=\cos\psi_0/\sin^2 \psi_0=\sqrt{t}/s$. Still need to invert $\psi_0=\psi(\tau_0)$ which yields $\tau_0=\sqrt{\frac{s}{2-s}}\mathbf{K}$.

There are two interesting critical values of $\tau_0$ around which we will explore the behavior of the correlator: $\tau_0^{\rm crit (1)}$ which corresponds to the value where the connected contribution becomes less important than the disconnected contribution and $\tau_0^{\rm crit(2)}$ which corresponds to the value of $\tau_0$ beyond which the connected solution ceases to exist. The concrete values are: 
\begin{equation}
   \tau_0^{\rm crit(1)}= 1.218, \quad  \tau_0^{\rm crit(2)}=1.875
   \quad \Leftrightarrow
\quad s^{\rm crit(1)}= 0.524, \quad  s^{\rm crit(2)}=0.790
\end{equation}
We verified that these values agree, in the appropriate limits, with those given originally in  \cite{Zarembo:1999bu} and more recently in \cite{Correa:2018lyl}.

The first sanity check consists in the correlators of insertions in the same loop to approximate the conformal limit when the separation distance vanishes. One can check that $\theta_1(z,q)$ for small $z$ goes linear in $z$. After taking two derivatives of $\log z $, we obtain the expected $\frac{1}{z^2}$ behavior for the correlators in the conformal limit with $\Delta=1$, as expected. This is seen in the plot in Fig. \ref{plotLL}, the dotted line described the conformal propagator. One can clearly see that for $\phi\to 0, 2\pi$, all correlators approximate the conformal one.

Figure \ref{plotLR} represents the two-point correlator of insertions on different loops. Its most salient feature is that it goes to a constant for small values of the separation $\phi$.

\begin{figure}[ht!]
\centering
\begin{tikzpicture}
\node at (-5.45,-0.85) {\includegraphics[width=7cm]{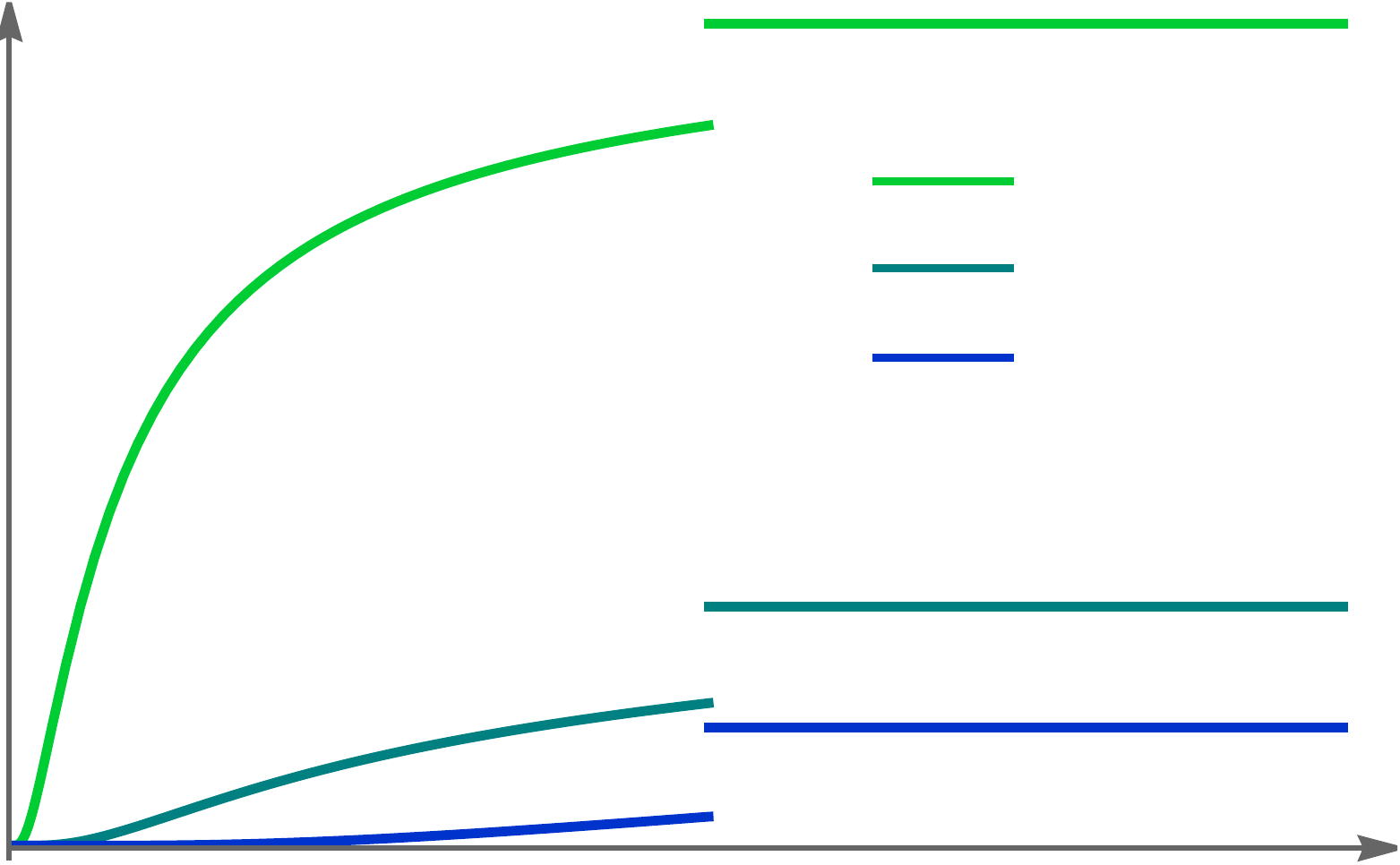}};
\draw[line width=0.3mm, gray] (-2.25,-2.96) -- (-2.25,-3.);
\node at (-2.25,-3.25) {$1$};
\node at (-1.7,-3.) {$s$};
\node at (-8.9,-3.25) {$0$};
\draw[line width=0.3mm, gray] (-5.4,-2.96) -- (-5.4,-3.);
\node at (-5.4,-3.25) {$0.524$};
\node at (-9.,1.7) {\small $\langle\mathcal{O}^{L}(\phi)\mathcal{O}^{L}(0)\rangle$ };
\node at (-3.25,0.45) {\small $\phi = \tfrac{\pi}{4}$};
\node at (-3.25,0.0) {\small $\phi = \tfrac{\pi}{2}$};
\node at (-3.25,-0.45) {\small $\phi = \pi$};
\end{tikzpicture}
\caption{$\langle\mathcal{O}^{L}(\phi)\mathcal{O}^{L}(0)\rangle$ correlator for fixed values of $\phi$, as a function of $s$. For $s>s^{{\rm crit}(1)}=0.524$ we depict the conformal correlator. }
\label{plotLL2}
\end{figure}
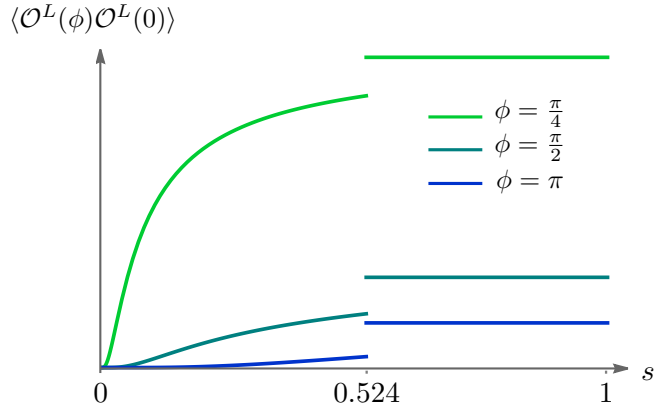

\begin{figure}[ht!]
\centering
\begin{tikzpicture}
\node at (-5.45,-0.85) {\includegraphics[width=7cm]{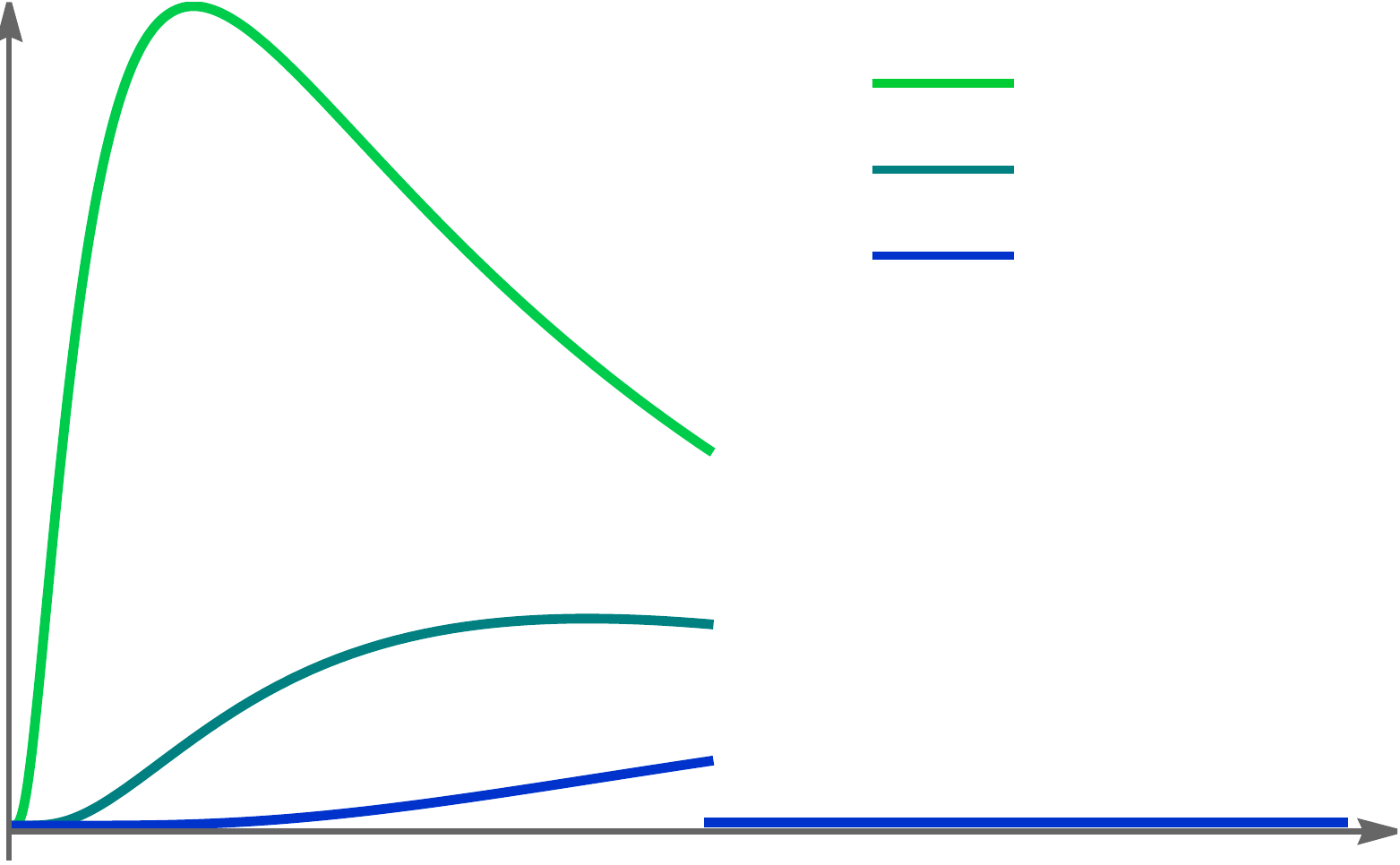}};
\draw[line width=0.3mm, gray] (-2.25,-2.96) -- (-2.25,-3.);
\node at (-2.25,-3.25) {$1$};
\node at (-1.7,-3.) {$s$};
\node at (-8.9,-3.25) {$0$};
\draw[line width=0.3mm, gray] (-5.4,-2.96) -- (-5.4,-3.);
\node at (-5.4,-3.25) {$0.524$};
\node at (-9.,1.7) {\small $\langle\mathcal{O}^{L}(\phi)\mathcal{O}^{R}(0)\rangle$ };
\node at (-3.25,0.9) {\small $\phi = \tfrac{\pi}{4}$};
\node at (-3.25,0.45) {\small $\phi = \tfrac{\pi}{2}$};
\node at (-3.25,0.) {\small $\phi = \pi$};
\end{tikzpicture}
\caption{$\langle\mathcal{O}^{L}(\phi)\mathcal{O}^{R}(0)\rangle$ correlator for fixed values of $\phi$, as a function of $s$.  For $s>s^{{\rm crit}(1)}=0.524$ the correlators are vanishing for all angles.}
\label{plotLR2}
\end{figure}
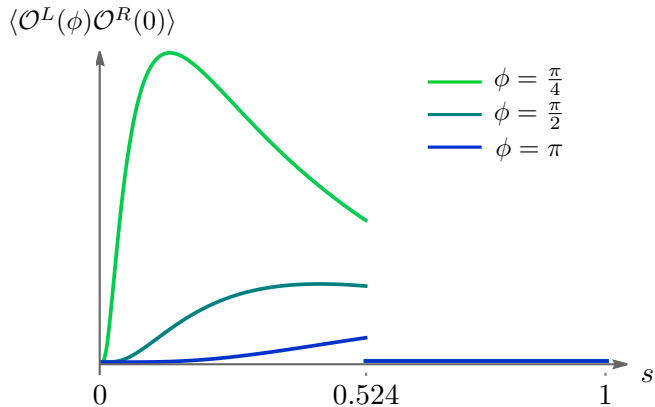

To the naked eye, the two-point functions are completely unremarkable, they appear as smooth continuous deformations of the conformal correlators. To demonstrate that such a point of view is too simplistic we plot in Figures \ref{plotLL2} and  \ref{plotLR2} the two-point correlators for a fixed value of the separation $\phi$ but as a function of $s$. The most salient feature is the gap between the solutions and the conformal correlator. In both cases this is quite pronounced and should clarify that the behavior is certainly different from the one of the conformal correlator. Note that we increase the value of the parameter $s$  to its critical value $s^{{\rm crit} (1)}= 0.524$, after which the disconnected world-sheet is dominant. Another noticeable feature is the vanishing of the correlators for finite $\phi$ as we take $s\to 0$. Essentially, this is the result of the bulk effectively disappearing, which prevents a perturbation in the boundary to propagate through the bulk to another point at some distance. The correlators simply become delta functions.

There is one extra feature in the Left-Right correlator as a function of $s$ that we now address: The appearance of a peak for small values of $\phi$ as we increase $s$, see Fig \ref{plotLR2}. We clarified that for $s\to 0$ the  correlator vanishes, it rises as we increase $s$. However, the Left-Right correlator must also become weaker again for longer world-sheets because the distance between the loops increases. The position of the peak depends on $\phi$. As can be seen in Fig. \ref{plotLR2}, smaller $\phi$ have a smaller value of $s$ of the peak. As we consider larger values of $\phi$, eventually the peak disapperas from the plot (or rather it goes beyond the critical value of $s^{{\rm crit}(1)}$). 

\subsubsection{Generic massless scalar}
In this subsection, we consider the scalars $\chi^{6,7,8,9}$ for general $K$. They satisfy \eqref{fe:eq5}, which has the general solution

\begin{equation}
    \label{fe:massless.sol.gen}
    \chi(\tau,\phi) = \sum\limits_{n=-\infty}^{\infty}\left[A_n\cosh\left(\omega_n\tau\right)+B_n\sinh\left(\omega_n\tau\right)\right]\e{in\phi}~,
\end{equation}
where 
\begin{equation}
    \label{fe:omega.n.def}
        \omega_n = \sqrt{n^2 -K^2}~.
\end{equation}
It is useful to set $A_n=0$ and switch off the source at one of he boundaries. The resulting correlators are 
\begin{align}
    \langle \mathcal{O}^{L,R}(\phi) \mathcal{O}^{L,R}(0) \rangle =& 
    -\sum\limits_{n=-\infty}^\infty \omega_n \coth(2\tau_0\omega_n)\e{i n \phi}~,
    \label{LLgenericK}
    \\
    \langle \mathcal{O}^{L,R}(\phi) \mathcal{O}^{R,L}(0) \rangle =& 
    \sum\limits_{n=-\infty}^\infty \frac{\omega_n}{\sinh(2\tau_0\omega_n)}\e{i n \phi}~.
    \label{LRgenericK}
\end{align}
The above expressions need to be slightly modified depending on the precise value of $K$, because the frequencies $\omega_n$ for $n^2<K^2$ are imaginary. 


\subsection{Massive scalar}

The scalar fields $\chi^{4,5}$ satisfy the field equation \eqref{fe:eq3},
\begin{equation}
\label{Eq:eq.massive.tau}
    \left(\partial_\tau^2 +\partial_\phi^2 - 2\cot^2 \psi + K^2 \right)\chi =0~,
\end{equation}
where we recall the relations 
$$ K^2 = \frac{1-s-t}{s}~,\qquad  s = \sin^2 \psi_0~,\qquad  t= C^2 \sin^4 \psi_0~, $$
from subsection \ref{bg:sec.on.shell.action}. In the nomenclature of subsection~\ref{ads.cft}, these fields have masses $m^2 L^2 = 2$, which implies $\Delta =2$ for the dual operators. 

First, introduce the Fourier modes $\chi_n$, for which \eqref{Eq:eq.massive.tau} becomes
\begin{equation}
\label{Eq:eq.massive.tau2}
    \left(\partial_\tau^2 - 2\cot^2 \psi + K^2  -n^2 \right)\chi_n =0~.
\end{equation}
We may view \eqref{Eq:eq.massive.tau2} as a Schr\"odinger equation with a periodic potential. With hindsight, let us proceed as follows. We introduce a lattice $\mathbb{L}=\{m+n\tilde{\tau}; m,n\in \mathbb{Z}\}$ with $\tilde{\tau}\in \mathbb{C}, \Im \tilde{\tau}>0$. The complex parameter $\tilde{\tau}$ will be determined later. 
Then, with 
\begin{equation}
\label{Eq:omegas}
    \omega_1=\frac12,\qquad 
    \omega_3 =\frac12 \tilde{\tau}~,\qquad
    \omega_2 =- \frac{1+\tilde{\tau}}{2}~,
\end{equation}
let $\wp(x)$ be the Weierstrass elliptic function with periodicity on $\mathbb{L}$. The roots are defined as usual by \cite{NIST:DLMF}
$$ e_i = \wp(\omega_i)~.$$
Now, let us perform the following variable transformation,
\begin{equation}
\label{Eq:tau.to.x}
    \frac{\tan^2\psi(\tau)}{\tan^2\psi_0} = \frac{e_1-e_2}{\wp(x)-e_2}~.
\end{equation}
The correspondence between particular coordinate values is illustrated in the table below. 
$$
\begin{array}{c|ccc}
\tau & 0 & \tau_0 & 2\tau_0 \\ \hline
x & 0 & \frac12 & 1
\end{array}
$$
With \eqref{Eq:tau.to.x}, \eqref{Eq:eq.massive.tau2} transforms into the Schr\"odinger equation\footnote{Originally, our path to \eqref{Eq:eq.massive.x} began with the change of variables $y=\frac{\tan^2\psi(\tau)}{\tan^2\psi_0}$ together with $\tilde{\chi}_n= y^{\frac12}\chi_n$, which gives rise to a Heun equation. Then, this Heun equation is transformed into \eqref{Eq:eq.massive.x} using the techniques of \cite{Takemura:2008aa}. It is, however, very easy to apply the transformation \eqref{Eq:tau.to.x} directly, which shows that $\tau = \sqrt{e_1-e_2} \tan\psi_0\, x$. This implies, in particular, $\tau_0 = \sqrt{e_1-e_2}\frac{\sqrt{s}}{2\sqrt{1-s}}$.\label{fn.vars}}
\begin{equation}
\label{Eq:eq.massive.x}
    \left[\partial_x^2 -2 \wp(x) + E_n \right]\chi_n =0~,
\end{equation}
where the roots satisfy the relation
\begin{equation}
\label{Eq:roots.rel}
    \frac{e_2-e_3}{e_1-e_2} = \frac{s+t}{1-s}
\end{equation}
and the energies $E_n$ are given by
\begin{equation}
\label{Eq:En}
    E_n = \frac{s(e_1-e_2)(1-n^2)}{(1-s)} -e_2~.
\end{equation}
The relation \eqref{Eq:roots.rel} determines the roots and the complex parameter $\tilde{\tau}$ via the standard relations 
\begin{equation}
\label{Eq:mod.rels}
    k^2 = \frac{e_2-e_3}{e_1-e_3}~, \qquad \omega_1 = \frac{\mathbf{K}}{\sqrt{e_1-e_3}}~,\qquad \omega_3 = i\frac{\mathbf{K}'}{\sqrt{e_1-e_3}}~.
\end{equation}
One easily finds that the modulus $k$ defined in \eqref{Eq:mod.rels} is just the modulus \eqref{bg:modulus} defined for the integrals in the background solution. Therefore, given the background parameters $s$ and $t$, from which one gets the modulus $k$ from \eqref{bg:modulus}, one calculates the lattice parameter $\tilde{\tau}$ and the roots from \eqref{Eq:mod.rels} and \eqref{Eq:omegas}. Specifically, one gets 
\begin{equation}
\label{Eq:roots.expl}
    e_3 = -\frac43 \mathbf{K}^2 (1+k^2)~,\qquad 
    e_1 = \frac43\mathbf{K}^2 (2-k^2)~,\qquad 
    e_2 = \frac43\mathbf{K}^2 (2k^2-1)~.
\end{equation}
Moreover, from \eqref{bg:dtheta.final} (or from the relation at the end of footnote \ref{fn.vars}) one finds the relation
\begin{equation}
    \tau_0 = \sqrt{\frac{s}{1+t}} \mathbf{K}~.
\end{equation}

Equation \eqref{Eq:eq.massive.x} represents the case of example 2.2 of \cite{Takemura:2008aa}. Taking the solutions from \cite{Takemura:2008aa}, one can easily construct the solution that has no source on the boundary at $x=0$. It is, up to an insignificant constant,
\begin{equation}
\label{Eq:eq.x.sol}
    \chi_n(x) = \sqrt{\wp(x)+E_n} \sinh\left[ \sqrt{-Q(E_n)}\int\limits_0^x \frac{\rmd x}{\wp(x)+E_n}\right]~, 
\end{equation}
where $Q(E)$ is the cubic polynomial 
\begin{equation}
\label{Eq:Q.def}
    Q(E) = (E+e_1)(E+e_2)(E+e_3)~. 
\end{equation}
The solution \eqref{Eq:eq.x.sol} is valid for $n^2\neq1$, because $E_{\pm1}=-e_2$, so that $Q(E_{\pm1})=0$. We shall consider the special case $n^2=1$ further below. 

In the general case, it remains to expand $\chi_n$ around $x=0$ and $x=1$ to read off the source and response coefficients, from which we will get the two-point functions. Close to $x=0$, we have
\begin{equation}
\label{Eq:x.expand.0}
    \chi_n(x) \approx \frac13 \sqrt{-Q(E_n)} x^2~,
\end{equation}
confirming that there is no source. Close to $x=1$, one gets
\begin{equation}
\label{Eq:x.expand.1}
    \chi_n(x) \approx \frac{\sinh \zeta_n}{1-x} - \frac13 \sqrt{-Q(E_n)} \cosh \zeta_n(1-x)^2~,
\end{equation}
where we have introduced the constants
\begin{equation}
\label{Eq:zeta.n.def}
    \zeta_n = \sqrt{-Q(E_n)} \int\limits_0^1 \frac{\rmd x}{\wp(x)+E_n}~.
\end{equation}
The integral can be converted to complete elliptic integrals. First, by symmetry,
$$
    \zeta_n = 2\sqrt{-Q(E_n)} \int\limits_0^{\frac12} \frac{\rmd x}{\wp(x)+E_n}~,
$$
after which we apply the change of integration variable
$$
1-y = \frac{e_1-e_2}{\wp(x)-e_2}~. 
$$
This yields 
\begin{align}
\notag
    \zeta_n &= \frac{2(1-s)\sqrt{-Q(E_n)}}{\sqrt{e_1-e_3}(e_1-e_2)} \int\limits_0^1 \frac{\rmd y\, \sqrt{1-y}}{\sqrt{y(1-k^2y)}[1-s+(1-n^2s)(1-y)]}\\
\notag &= 
    \frac{2(1-s)\sqrt{-Q(E_n)}}{\sqrt{e_1-e_3}(e_1-e_2)s(1-n^2)} 
    \left[ \mathbf{K} -\frac{1-s}{1-n^2s} \mathbf{\Pi}\left(\frac{(1-n^2)s}{1-n^2s}\right)  \right]~.
\end{align}

Let us also write out the polynomial $Q(E_n)$ using \eqref{Eq:Q.def}, \eqref{Eq:En}, and \eqref{bg:modulus}, 
\begin{equation}
    \label{Eq:Q.result}
    Q(E_n) = - \frac{s(1-n^2)(1-n^2s)(t+n^2s)}{(1+t)(1-s)^2}(e_1-e_3)(e_1-e_2)^2~.
\end{equation}
Thus, one obtains (recall $n\neq \pm1$)
\begin{equation}
\label{Eq:zeta.result}
    \zeta_n = 2 \sqrt{\frac{(1-n^2s)(t+n^2s)}{s(1+t)(1-n^2)}} 
    \left[ \mathbf{K} -\frac{1-s}{1-n^2s} \mathbf{\Pi}\left(\frac{(1-n^2)s}{1-n^2s}\right)  \right]~.
\end{equation}
It is an amazing coincidence that $\zeta_0$ is
\begin{equation}
\label{Eq:zeta.0}
    \zeta_0 = 2 \sqrt{\frac{t}{s(1+t)}} 
    \left[ \mathbf{K} - (1-s) \mathbf{\Pi}(s)  \right] =2J~,
\end{equation}
with $J$ being the background geometry parameter \eqref{bg:J.def}, as can be seen from \eqref{bg:alpha.final} and \eqref{bg:modulus}. 

In the special case $n^2=1$, the solution of \eqref{Eq:eq.massive.x} without source at the boundary at $x=0$ is, up to an irrelevant constant,  
\begin{equation}
\label{Eq:eq.x.sol.spec}
    \chi_{\pm1}(x) = \sqrt{\wp(x)-e_2} \int\limits_0^x \frac{\rmd x}{\wp(x)-e_2}~. 
\end{equation}
Expanding it close to the boundaries at $x=0$ and $x=1$ yields 
\begin{align}
    \chi_{\pm1}(x) \approx \frac13 x^2~, \qquad \chi_{\pm1}(x) \approx \frac{\tilde{\zeta}}{1-x} -\frac13 (1-x)^2~,
\end{align}
respectively, where the constant $\tilde{\zeta}$ is 
\begin{equation}
\label{Eq:zeta.1}
    \tilde{\zeta} = \int\limits_0^1 \frac{\rmd x}{\wp(x)-e_2} = 
    \frac{\mathbf{E} - (1-k^2) \mathbf{K}}{4k^2(1-k^2)\mathbf{K}^3} = -\frac1{8k} \frac{\rmd}{\rmd k} \mathbf{K}^{-2}~.
\end{equation}
The integral can be done using the same transformation as in the general case.
Again, it is an amazing coincidence that $\tilde{\zeta}$ is proportional to the on-shell action \eqref{bg:S.ren},
\begin{equation}
\label{Eq:zeta.1.S}
    \tilde{\zeta} = -\frac{S_{ren}}{\sqrt{\lambda}} \frac{s^2}{(s+t)(1-s)(2\tau_0)^3}~.
\end{equation}

Finally, we can write down the two-point functions for the operators that are dual to the scalars $\chi^{4,5}$. We separate the modes $n=-1,0,1$, for which we have the special relations \eqref{Eq:zeta.0} and {\eqref{Eq:zeta.1.S}}, as well as 
\begin{equation}
    \label{Eq:Q.0}
    \sqrt{-Q(E_0)} = \frac{e_1-e_2}{1-s} \sqrt{\frac{st}{1+t}(e_1-e_3)} = 8 \mathbf{K}^3 \sqrt{\frac{st}{(1+t)^3}} = \frac{(2\tau_0)^3\sqrt{t}}{s}~. 
\end{equation}
Similarly, we write from \eqref{Eq:Q.result}
\begin{equation}
    \label{Eq:Q.n}
    \sqrt{-Q(E_n)} = \frac{(2\tau_0)^3}{s}\sqrt{(1-n^2)(1-n^2s)(t+n^2s)}~. 
\end{equation}
The factor $(2\tau_0)^3$ compensates the variable re-scaling, as discussed at the end of subsection~\ref{ads.cft}.

Therefore, we obtain the two-point functions
\begin{align}
    \langle\mathcal{O}^{L,R}(\phi)\mathcal{O}^{L,R}(0) \rangle &= -\frac{\sqrt{t}}{s\sqrt{1-\alpha^2}} +\cos\phi\frac{2\sqrt{\lambda}(s+t)(1-s)}{s^2 S_{ren}} \\
\notag
    &\quad 
    -\frac{2}{s} \sum\limits_{n=2}^{\infty} \cos(n\phi) \sqrt{(1-n^2)(1-n^2s)(t+n^2s)} \coth \zeta_n~,\\
    \langle\mathcal{O}^{L,R}(\phi)\mathcal{O}^{R,L}(0) \rangle &= 
    \frac{\sqrt{t}\alpha}{s\sqrt{1-\alpha^2}} -\cos\phi\frac{2\sqrt{\lambda}(s+t)(1-s)}{s^2 S_{ren}} \\
\notag
    &\quad     
    + \frac{2}{s} \sum\limits_{n=2}^{\infty} \cos(n\phi) \frac{\sqrt{(1-n^2)(1-n^2s)(t+n^2s)}}{\sinh \zeta_n}~.  
\end{align}
Here, $\alpha$ is the geometric invariant parameter \eqref{conf.invar}, as follows from \eqref{bg:alpha}. Let us briefly make two comments about the above results. For $n^2>1$, we would need $n^2s>1$ in order for the square root to be real. But also $\zeta_n$ contains such a square root, so the imaginary units cancel out, after the hyperbolic functions have been converted to trigonometric ones. Thus, there is no problem with the reality of the two-point functions. Note also that the $\sinh$ in the denominators has zeros at $\zeta_n=im\pi$, which indicate resonances. At the moment we are not quite sure how to interpret this possibility.

\subsection{Geodesic approximation for heavy scalars}
Thus far, we have rigorously described the light operators as fluctuations of the string world-sheet. In this section we take a more phenomenological approach and consider the correlators of very heavy operators. These operators might arise, for example, as very massive string states. One expects that for such operators the correlator can be well approximated by $\langle {\cal O}(\phi_1) {\cal O}(\phi_2)\rangle \sim \exp\left(-m_{\cal O} \,\ell(\phi_1,\phi_2)\right)$, where $\ell(\phi_1,\phi_2)$ is the (renormalized) length of the \emph{shortest} geodesic connecting the two insertion points of the operators, which might be located on the same or on different boundaries of the world-sheet.

An example of the operators we have in mind were dubbed ``two-particle operators'' in  \cite{Giombi:2017cqn}. Consider an operator of the form  $\Phi^a \partial_\phi^{2n}\Phi^b$ where one usually takes the singlet, symmetric traceless, or antisymmetric representations in the $SO(5)$ indices $a,b$. Such operators appear in the OPE expansion computing the four-point function of simple operator insertions such as those corresponding to fields discussed in the previous subsections. An impressive result of \cite{Giombi:2017cqn} was to determine corrections to the conformal dimension:  $\Delta=2+2n-(2n^2+3n)/\sqrt{\lambda}$. Our world-sheet configuration is more general than the dual of the $\frac{1}{2}$-BPS Wilson-Maldacena loop considered in \cite{Giombi:2017cqn}. In particular, we do not have a natural organization in terms of irreducible representations of $SO(5)$, although $SO(4)$ seems to survive. We use some of the supersymmetric results as guidance for our current non-supersymmetric, non-conformal situation.

The geodesic analysis relevant for such heavy operators is as follows. 
We start with the expression of the geodesic length, 
\begin{equation}\label{Eq:geodesic.length}
    \ell = L \int \rmd u\, \cot\psi \sqrt{\dot{\tau}^2 +\dot{\phi}^2}~,
\end{equation}
where a dot represents a derivative with respect to $u$. The equation of motion for $\sigma$ is solved by\footnote{The equation of motion for $\tau$ is then implied by \eqref{Eq:geodesic:sigma} because of diffeomorphism invariance.}
\begin{equation}\label{Eq:geodesic:sigma}
    \dot{\phi} = P \tan \psi  \sqrt{\dot{\tau}^2 +\dot{\phi}^2}~,
\end{equation}
where $P$ is an integration constant. Without loss of generality, we can set $u=\sigma$ and obtain
\begin{equation}\label{Eq:geodesic:tau}
    \rmd \phi = \pm \frac{\tan\psi\, \rmd \tau}{\sqrt{\tan^2\psi_P -\tan^2\psi}}~,
\end{equation}
where we have defined $\tan^2\psi_P=P^{-2}$. To continue, we write $\rmd \tau = \rmd \psi/\psi'$ using \eqref{bg:diff.new} and change variable by defining $y = \frac{\tan^2 \psi}{\tan^2\psi_0}$, with $\psi_0$ defined in \eqref{bg:K.sub}. Using also \eqref{bg:s.t.def}, this  results in
\begin{equation}\label{Eq:geodesic:sigma.u}
    \rmd \phi = \pm  \frac{\frac12 \sqrt{\frac{s}{1-s}}\rmd y}{\sqrt{(y_P-y)(1-y)\left( 1+\frac{s+t}{1-s}y\right)}}~,
\end{equation}
where we have defined the new constant $y_P=\frac{\tan^2\psi_P}{\tan^2\psi_0}$. 

We must distinguish two cases. For $y_P<1$, the geodesic reaches a certain $\tau=\tau_P<\tau_0$ and then turns back to the boundary it started at. These geodesics represent the Left-Left (or Right-Right) correlators. 
For $y_P>1$, the geodesic reaches the middle of the world-sheet at $\tau=\tau_0$, and then continues to the other boundary. These represent the Left-Right correlators. In either case, the difference $\phi=\phi_2-\phi_1$ is given by
\begin{equation}\label{Eq:geodesic:phi}
   \phi =  \sqrt{\frac{s}{1-s}} \int\limits_0^{\min(y_P,1)} \frac{\rmd y}{\sqrt{(y_P-y)(1-y)\left( 1+\frac{s+t}{1-s}y\right)}}~.
\end{equation}
Clearly, the integral on the right hand side diverges in the limiting case $y_P=1$. Because $\phi$ counts only modulus $2\pi$, the fact that $\phi$ can be arbitrarily large implies that there are infinitely many geodesics connecting any two boundary points. They can be distinguished by their ``winding number'' $w$, which measures how many times and in which sense they wind around the world-sheet. We are interested only in the shortest of these, because it dominates the amplitude $\e{-m_{\cal O}\ell}$.\footnote{In principle, one should write $\sum_w \e{-m_{\cal O}\ell_w}$ with $\phi_w=\phi+2\pi w$.}

The geodesic length can be found using the same manipulations as above. One finds 
\begin{equation}\label{Eq:geodesic:l}
   \ell =  L \int\limits_\epsilon^{\min(y_P,1)} \frac{\sqrt{y_P}\rmd y}{y\sqrt{(y_P-y)(1-y)\left( 1+\frac{s+t}{1-s}y\right)}}~,
\end{equation}
where we have introduced the cut-off $\epsilon$ to regulate the logarithmic divergence at the world-sheet boundary. One must renormalize this by subtracting the universal divergence as described in \cite{Aharony:1999ti}. For example, one may subtract the value in the limit $y_P\to\infty$, which is independent of $y_P$ and represents the length of a geodesic stretching between the two boundaries with $\phi=0$. Alternatively, one can add a boundary term that implements the appropriate boundary conditions and regulates the divergence \cite{Drukker:1999zq}. The integrals themselves can be done and result in elliptic integrals, but they are not very illuminating.

For small $y_P$, when the geodesic stays close to the boundary, after changing variable $y=y_Pz$, $\ell$ diverges as $\ln\frac{y_P}{\epsilon}$, which becomes $\ln\frac{y_P}{\mu}$ with some renormalization scale $\mu$ after subtracting the universal divergence. On the other hand, we get from \eqref{Eq:geodesic:phi} $\phi \sim \sqrt{y_P}$, so that the leading amplitude becomes $\e{-m_{\cal O} \ell} \sim \phi^{-2m_{\cal O}L}$, reproducing the conformal behaviour.

\section{Field theory correlators  in the ladder approximation}\label{Sec:FTside}

Field theory computations in the context of the half-BPS Maldacena-Wilson loop have a rich history with some important recurrent themes that we briefly review to motivate the structure we follow in this section. The fact that the combined propagator of the gluon and the scalar field in a circular Wilson loop is constant motivated the Gaussian matrix model conjecture \cite{Drukker:2000rr,Erickson:2000af}. The latter states that expectation values are computed using a Gaussian matrix model which in practice resums  ladder diagrams; this conjecture was later proven by Pestun  \cite{Pestun:2007rz}. 
The combined propagator for insertions in two different loops is not constant, as will be shown in Equation \eqref{G(theta)}, and the ladder diagrams do not provide the complete perturbative description of the problem. Computing the resummation of ladder diagrams can be nevertheless very instructive.
Summing ladders is a venerated tradition in this context, not only it is at the heart of the original conjecture stating that the vacuum expectation value of the half-supersymmetric Wilson loop was determined by a Gaussian matrix model \cite{Drukker:2000rr,Erickson:2000af} but it also  can be used to extract a qualitatively correct picture of the strong coupling description. For instance, the resummation of ladders diagrams for the correlator of two circular Wilson loops exhibits, in the strong coupling limit, a phase transition quite similar to the Gross-Ooguri one \cite{Zarembo:2001jp, Correa:2018pfn}. Additionally, the ladder truncation can also be justified when certain analytic continuation of the coupling of the Wilson loops to the scalar fields is considered. This has allowed some explicit connection between the resummation of ladder and string theory results \cite{Correa:2012at,Correa:2018lyl}. 

With these motivations in mind, in this section we concentrate on the ladder diagrams contribution to the correlator of insertions in two Wilson loops. Our analysis will closely follow that of \cite{Zarembo:2001jp, Correa:2018pfn}, where the ladder contribution to the expectation value of connected correlators of Wilson-Maldacena loops are given in terms of certain  Green's functions satisfying a set of  Dyson equations.
We consider inserting both operators in the same  loop and also one operator per loop. {\it Our main finding is the fact that the very same Green's functions that describe the expectation value of the correlator of two Wilson-Maldacena  loops also describe the correlation between insertions.}

\subsection{The correlator of two Wilson loops}

We study correlations between excitations inserted along Wilson loops in the ladder approximation, {\it i.e.}, adding up only  Feynman diagrams with no interaction vertices. In principle, we will consider this approximation in the large $N$ limit and for arbitrary values of the 't Hooft coupling $\lambda = g_{_{  Y\!M}}^2N$.
Admittedly, this turns out to be an uncontrolled approximation in the general case. However, as we shall see in section \ref{largecos}, this procedure becomes a sound approximation to the leading contribution in a certain parametric limit. More concretely, when taking the limit
$\cos\gamma \to \infty$ while keeping $\lambda\cos\gamma$
fixed  ladder diagrams become dominant,
as diagrams with interacting vertices become suppressed by powers of $1/\cos\gamma$. This kind of limit was first considered for a Wilson loop with a cusp angle in \cite{Correa:2012nk}  and later on extended for correlators of Wilson loops in \cite{Correa:2018pfn}.   

Consider the two Wilson-Maldacena loops defined in \eqref{WLdef} with the contours \eqref{C12}.
Let us first review the evaluation of the vacuum expectation value of the above two Wilson-Maldacena loops as we introduce some convenient technical ingredients, more details can be found in  \cite{Zarembo:2001jp,Correa:2018pfn}.
To compute ladder diagrams, it is convenient to define the following Gaussian effective fields 
\begin{equation}
\varphi_{a}(\phi) \equiv g_{_{  Y\!M}} \left(i A_\mu\dot{x_a}^\mu  + \Phi _I n_a^I |\dot{x_a}|\right)
\end{equation}
where $a=1,2$ indicates on which loop do they live. The propagator of these effective fields are \cite{Drukker:2000rr,Erickson:2000af}:
\begin{equation}
 \left\langle \varphi^i_{a\,j}(\phi) \varphi^k_{b\,l}(\phi')\right\rangle=
 \frac{1}{N}\,\delta ^i_l\delta ^k_jG_{ab}(\phi-\phi')~,
\end{equation}
where
\begin{equation}
 G_{11}=G_{22}=\frac{\lambda }{16\pi ^2}\equiv g~,
 \label{propa11}
\end{equation}
while
\begin{equation}\label{G(theta)}
 G_{12}(\phi )=G_{21}(\phi )=g \,\frac{\cos\gamma +\cos \phi }{\alpha^{-1}-\cos \phi }\equiv G(\phi )~,
\end{equation}
with $\alpha$ denoting the conformally invariant geometric parameter defined in \eqref{conf.invar}, see also \eqref{bg:alpha}.
In a sense, restricting the summation of diagrams to ladder ones is the same as treating the theory as if it were Gaussian, as we compute expectation values using free field theory propagators only.\footnote{The non-trivial $g$-dependence that appears in the expressions below arises from an explicit $g_{_{  Y\!M}}$ dependence in the Wilson \eqref{WLdef}. }

%

It is convenient to introduce the following auxiliary operators describing two path-ordered exponentials: 
\begin{eqnarray}
 && \PU{\!_a}(\phi_1,\phi_2)=\vv{P}\exp\int_{\phi_1}^{\phi_2}d\phi\,
 \varphi_a(\phi),
 \qquad
\AU_{\!a}(\phi_1,\phi_2)=\cvv{P}\exp\int_{\phi_1}^{\phi_2}d\phi\,
   \varphi_a(\phi),
  \label{AU-recu}
\end{eqnarray}
where $a=1,2$ indicates the circular loop that is being considered. The symbols $\vv{P}$ and $\cvv{P}$ denote path and anti-path ordering: the rightmost field in the expansion of the exponential has the largest, respectively, the smallest argument. The Wilson loops can be defined in terms of these operators.
Writing the ordered exponentials as
\begin{equation}
 \PU_{\!a}(\phi_1,\phi_2) = \prod_{\phi\in (\phi_1,\phi_2)
}\left(\mathbbm{1} + \varphi_a(\phi) d\phi\right),
\end{equation}
it is straightforward to verify that they satisfy the following recursive relations
\begin{eqnarray}
\PU_{\!a}(\phi_1,\phi_2) =\mathbbm{1}+\!\int_{\phi_1}^{\phi_2}\!\!\!d\phi\,\PU_{\!a}(\phi_1,\phi)\varphi_a(\phi),
\label{AU-rec} 
\quad
\AU_{\!a}(\phi_1,\phi_2) =\mathbbm{1}+\!\int_{\phi_1}^{\phi_2}\!\!\!d\phi\,\varphi_a(\phi)\AU_{\!a}(\phi,\phi_2).
\end{eqnarray}

Consider for instance
\begin{equation}\label{defW}
 W(\phi) =\frac{1}{N} \left\langle\mathop{\mathrm{tr}}\AU_{\!1}(0,\phi)\right\rangle.
\end{equation}
Using \eqref{AU-rec}, the Wick's theorem and large-$N$ factorization 
we get a simple Dyson equation
\begin{equation}\label{DysonW}
W(\phi)=1+g\int_{0}^{\phi}d\phi'\int_{0}^{\phi'}d\phi''\,W(\phi'-\phi'')W(\phi'')\,.
\end{equation}
This equation for $W(\phi)$, which involves only the constant propagator (\ref{propa11}), can be easily solved in terms of the Laplace transform, $W(z)$,  defined according to
\begin{equation}
W(z)  = \int\limits^{\infty}_0 d\phi\, e^{-z \phi}\, W(\phi).
\end{equation}
Then,
\begin{equation}
W(z) = \frac{1}{z} +g\frac{W(z)^2}{z}\quad
\Rightarrow\quad
W(z) = \frac{z-\sqrt{z^2-4g}}{2g},
\end{equation}
and anti-transforming
\begin{equation}
W(\phi) = \frac1{\sqrt{g}\phi} I_1(2\sqrt{g}\phi),
\end{equation}
where $I_1$ is a modified Bessel function of the first kind. This function, evaluated at $2\pi$, gives the well-known result for the expectation value of the circular Wilson-Maldacena loop in the large $N$ limit \cite{Drukker:2000rr,Erickson:2000af}.  It is worth remarking the important role that a constant propagators plays in the road to certain results. We will highlight this threat in various computations and thus motivate certain limits in the space of parameters.
In fact, in a similar fashion, the connected correlator of $k$ traces of a unique Wilson loop 
\begin{equation}
  W(\phi_1,\cdots,\phi_k)= N^{k-2} 
\left\langle\mathop{\mathrm{tr}}\AU_{\!1}(0,\phi_1)\cdots\mathop{\mathrm{tr}}\AU_{\!1}(0,\phi_k)\right\rangle_{\rm conn},
\end{equation}
can also be explicitly computed in terms of constant propagators. For example, for $k=2$ one has \cite{Akemann:2001st}
\begin{equation}
    W(\phi_1,\phi_2) = \frac{\sqrt{g}\phi_1 \phi_2}{\phi_1 + \phi_2} \left[I_0(2\sqrt{g}\phi_1)I_1(2\sqrt{g}\phi_2)
    +I_1(2\sqrt{g}\phi_1)I_0(2\sqrt{g}\phi_2)
    \right]\,.
\end{equation}

Let us now turn to the correlator of the two concentric coaxial Wilson-Maldacena loops given in \eqref{C12}. This configuration is non-supersymmetric and its ladder contribution can be tackled with similar methods. 
It is convenient to introduce two Green's functions,
\begin{equation}
 K(\phi)=\left\langle \mathop{\mathrm{tr}}\AU_{\!1}(0,\phi)\mathop{\mathrm{tr}}\PU_{\!2}(0,2\pi )\right\rangle_{\rm conn},
\label{K}
\end{equation}
and
\begin{equation}\label{Gamma}
 \Gamma (\phi_1,\phi_2|\varphi )=\left\langle \frac{1}{N}\,\mathop{\mathrm{tr}}\AU_{\!1}(0,\phi_1)\PU_{\!2}(\varphi ,\varphi +\phi_2)\right\rangle.
\end{equation}
Both are quadratic in ordered exponentials $\AU_{\!1}$ and $\PU_{\!2}$, but differ in the number of traces. The ladder contribution to the connected correlator of the two Wilson loops is simply given by
\begin{equation}
 \left\langle W(C_1)W(C_2)\right\rangle_{\rm ladders}=K(2\pi).
\end{equation}
The derivation of a Dyson equation that relates the Green's function $K(\phi)$ and $\Gamma(\phi_1,\phi_2|\varphi)$ was given in \cite{Correa:2018pfn} and we review it here. Using
\eqref{AU-rec}, it follows that 
\begin{equation}
    \left\langle \mathop{\mathrm{tr}}\AU_{\!1}(0,\phi)\mathop{\mathrm{tr}}\PU_{\!2}(0,2\pi )\right\rangle
 = N^2 W(2\pi) +
 \int_{0}^{\phi}d\phi'\,
 \left\langle \mathop{\mathrm{tr}}
 \varphi_1(\phi') \AU_{\!1}(\phi',\phi)\mathop{\mathrm{tr}}\PU_{\!2}(0,2\pi )\right\rangle.
\end{equation}
The first term in the right hand side can be re-written using \eqref{DysonW}. There are two possible Wick's contractions for the second term, giving rise to propagators $G_{11}=G_{22}$ and $G_{12}=G_{21}$ appearing  in \eqref{propa11}-\eqref{G(theta)}, respectively,
\begin{align}
 \left\langle \mathop{\mathrm{tr}}\AU_{\!1}(0,\phi)\mathop{\mathrm{tr}}\PU_{\!2}
 (0,2\pi )\right\rangle
 = &\, N^2 W(2\pi)\left(
 W(\phi)-\int_{0}^{\phi}\!\!d\phi'\!\int_{0}^{\phi'}\!\!\!d\phi''\,
 W(\phi'-\phi'')W(\phi'')
 \right)
 \\ &
 +\frac{g}{N}\!\int_{0}^{\phi}\!d\phi'\!\int_{0}^{\phi'}\!d\phi''
 \left\langle \mathop{\mathrm{tr}}\AU_{\!1}(0,\phi'')\mathop{\mathrm{tr}}\AU_{\!1}(\phi'',\phi')\mathop{\mathrm{tr}}\PU_{\!2}(0,2\pi )\right\rangle
\nonumber \\ 
&
+\frac{1}{N}\!\int_{0}^{\phi}\!d\phi'\!\int_{0}^{2\pi }\!d\varphi \,G(\varphi -\phi')
\left\langle \mathop{\mathrm{tr}}\AU_{\!1}(0,\phi')\PU_{\!2}(\varphi ,\varphi +2\pi )\right\rangle.
\nonumber
\end{align}
Applying large-$N$ factorization and removing the disconnected part of the correlators we obtain an equation that relates $K$ to $\Gamma $,
\begin{equation}\label{DysonK}
K(\phi)=2g\int_{0}^{\phi}d\phi'\int_{0}^{\phi'}d\phi''\,W(\phi'-\phi'')K(\phi'')
 +\int_{0}^{\phi}d\phi'\int_{0}^{2\pi }d\varphi \,G(\varphi -\phi')\Gamma (\phi',2\pi |\varphi ).
\end{equation}

The Green's function $\Gamma (\phi_1,\phi_2|\varphi )$  satisfies a closed Dyson equation, which can be derived  following similar arguments
\begin{eqnarray}\label{DysonGamma}
 \Gamma (\phi_1,\phi_2|\varphi)&=&W(\phi_2)+g\int_{0}^{\phi_1\!}dt'\int_{0}^{\phi'}\!\!d\phi''\,W(\phi'-\phi'')\Gamma (\phi'',\phi_2|\varphi )
\nonumber \\
&&+\int_{0}^{\phi_1}\!d\phi'\int_{0}^{\phi_2}\!\!d\phi''\,G(\varphi +\phi''-\phi')W(\phi_2-\phi'')\Gamma (\phi',\phi''|\varphi ).
\end{eqnarray}

This equation can be brought to a more symmetric form. Eq. \eqref{DysonGamma}
is an integral equation of the type 
\begin{equation}
f(\phi)=g\int_{0}^{\phi}\!d\phi'\int_{0}^{\phi'}\!\!d\phi''\,W(\phi'-\phi'')f (\phi'')+\int_{0}^{\phi}\!d\phi'\,j(\phi').
 \label{identity1}
\end{equation}
Using of the Dyson equation (\ref{DysonW}) for $W(\phi)$ it is easy to see that this can be solved by
\begin{equation}
f(\phi)=\int_{0}^{\phi}\!d\phi'\,W(\phi-\phi')j(\phi').
  \label{identity2}
\end{equation}
Applying this result to the Dyson equation (\ref{DysonGamma}) brings the latter to a symmetric form
\begin{align}
\label{improvedDysonG}
 \Gamma (\phi_1,\phi_2|\varphi)= & W(\phi_1)W(\phi_2)
 \\
 &+\int_{0}^{\phi_1}d\phi'\int_{0}^{\phi_2}d\phi''\,
 W(\phi_1-\phi')W(\phi_2-\phi'')G(\varphi +\phi''-\phi')\Gamma (\phi',\phi''|\varphi ),
 \nonumber
\end{align}

These Dyson equations are in general difficult to solve. There exist, however, certain limits in which they simplify considerably. One limit of interest is $\alpha\to 1$ and $\cos\gamma \to -1$. In this limit the two loops with opposite orientations become coincident and supersymmetric. The effective propagator $G(\phi)$ becomes constant and Dyson equations \eqref{DysonK} and \eqref{improvedDysonG} can be simply solved via a Laplace transform, as done with the Green's function $W(t)$.

Another limit of interest is the analytic continuation  $\cos\gamma \to \infty$. In first place, because being $G(\phi)\gg g$ one can set $W(\phi)$ to 1 in Dyson equations \eqref{DysonK} and \eqref{improvedDysonG} to capture the leading order contribution in this parametric limit. Additionally, the non-ladder contributions are expected to
be suppressed in this limit.

\subsection{Inserting local operators}
With the preliminaries covered, we can now tackle the insertion of operators in the correlator of two concentric circular loops. We might either consider the insertions in the same Wilson loop or insertions at different loops. It is possible to compute the correlation function between these insertions through the following expectation values
\begin{equation}
\langle\!\langle {\cal O}_1^L(\phi_1) {\cal O}^L_2(\phi_2)\rangle\!\rangle =
\frac{
\langle {\rm tr}[P e^{i\oint_{C_1} d\phi \varphi_1(\phi)}{\cal O}_1(\phi_1){\cal O}_2(\phi_2)]
{\rm tr}[P e^{i\oint_{C_2}d\phi' \varphi_2(\phi')}]\rangle_{\rm conn}
}
{
\langle {\rm tr}[ P e^{i\oint_{C_1} d\phi \varphi_1(\phi)}]
{\rm tr}[Pe^{i\oint_{C_2}d\phi' \varphi_2(\phi')}]\rangle_{\rm conn}
},
\label{2ptLL}
\end{equation}
\begin{equation}
\langle\!\langle {\cal O}_1^L(\phi_1) {\cal O}^R_2(\phi_2)\rangle\!\rangle =
\frac{
\langle {\rm tr}[P e^{i\oint_{C_1} d\phi \varphi_1(\phi)}{\cal O}_1(\phi_1)]
{\rm tr}[P e^{i\oint_{C_2}d\phi' \varphi_2(\phi')}{\cal O}_2(\phi_2)]\rangle_{\rm conn}
}
{
\langle {\rm tr}[ P e^{i\oint_{C_1} d\phi \varphi_1(\phi)}]
{\rm tr}[Pe^{i\oint_{C_2}d\phi' \varphi_2(\phi')}]\rangle_{\rm conn}
},
\label{2ptLR}
\end{equation}
where, for the case of two traces, the connected part means
\begin{equation}
    \langle
    {\rm tr}(A) {\rm tr}(B)
    \rangle_{\rm conn} 
    = \langle
    {\rm tr}(A) {\rm tr}(B)
    \rangle - \langle
    {\rm tr}(A)\rangle  \langle{\rm tr}(B)
    \rangle. 
\end{equation}
Armed with the intuition of the previous section, we would like to compute the ladder diagrams contributions to those expectation values. Among all candidates for operator insertions, we will restrict our attention to the simplest possibility: we will consider insertions of scalar fields that do not appear in the Wilson loops ($\Phi_1,\cdots \Phi_4$). For them, the types of ladder diagrams are limited and their holographic duals are easily identifiable on the string theory description.

\subsubsection{Insertions in the same loop}
Let us compute \eqref{2ptLL} in the ladder approximation
For the numerator of \eqref{2ptLL} we need
\begin{eqnarray}
&&\langle {\rm tr} \left(\AU_{\!1}(0,\phi_1)\Phi_1(\phi_1)\AU_{\!1}(\phi_1,\phi_2)\Phi_1(\phi_2)\AU_{\!1}(\phi_2,2\pi)\right)
{\rm tr} \left(\PU_{\!2}(0,2\pi)\right)\rangle
\nonumber
\\
&& = \Delta_{11}(\phi_2-\phi_1) \langle {\rm tr} \left(
\AU_{\!1}(\phi_2 ,2\pi+\phi_1)\right)  {\rm tr} \left( \AU_{\!1}(\phi_1,\phi_2)\right)
{\rm tr} \left(\PU_{\!2}(0,2\pi)\right)\rangle
\label{numLL1}
\end{eqnarray}
where the contraction between the two $\Phi_1$ splits one single trace into two  and introduces a propagator 
\begin{equation}
\Delta_{11}(\phi) = \frac{g}{1 -\cos \phi}\,.
\end{equation}
To arrive at the expression \eqref{numLL1} we used that
${\rm tr}\left(\AU_{\!1}(0,\phi_1)\AU_{\!1}(\phi_2,2\pi)\right) = {\rm tr}\left(\AU_{\!1}(\phi_2,2\pi+\phi_1)\right)$. In the large $N$ approximation, the vev of three traces can be expanded as follows
\begin{eqnarray}
 && \langle {\rm tr} \left(
\AU_{\!1}(\phi_2,2\pi+\phi_1)\right)  {\rm tr} \left( \AU_{\!1}(\phi_1,\phi_2)\right)
{\rm tr} \left(\PU_{\!2}(0,2\pi)\right)\rangle 
\nonumber \\
&&=
\langle {\rm tr} \left(
\AU_{\!1}(\phi_2,2\pi+\phi_1)\right)\rangle \langle  {\rm tr} \left( \AU_{\!1}(\phi_1,\phi_2)\right)\rangle \langle
{\rm tr} \left(\PU_{\!2}(0,2\pi)\right)\rangle 
\nonumber \\
&&+
\langle {\rm tr} \left(
\AU_{\!1}(\phi_2,2\pi+\phi_1)\right)\rangle \langle  {\rm tr} \left( \AU_{\!1}(\phi_1,\phi_2)\right)
{\rm tr} \left(\PU_{\!2}(0,2\pi)\right)\rangle_{\rm conn}
\nonumber \\
&&+
\langle {\rm tr}  \left( \AU_{\!1}(\phi_1,\phi_2)\right) \rangle
\langle  {\rm tr} \left(\AU_{\!1}(\phi_2 ,2\pi+\phi_1)\right)
{\rm tr} \left(\PU_{\!2}(0,2\pi)\right)\rangle_{\rm conn}
\nonumber \\
&&+
\langle {\rm tr}  \left(\PU_{\!2}(0,2\pi)\right) \rangle
\langle  {\rm tr} \left( \AU_{\!1}(\phi_1,\phi_2)\right) {\rm tr} \left(\AU_{\!1}(\phi_2,2\pi+\phi_1)\right)
\rangle_{\rm conn} 
\nonumber \\
&&+
\langle {\rm tr} \left(
\AU_{\!1}(\phi_2 -\phi_1,2\pi)\right)  {\rm tr} \left( \AU_{\!1}(\phi_1,\phi_2)\right)
{\rm tr} \left(\PU_{\!2}(0,2\pi)\right)\rangle_{\rm conn}.
\end{eqnarray}
The first line in the right hand side of this equation, $N^3 W(2\pi-\phi_2+\phi_1)W(\phi_2-\phi_1)W(2\pi)$, is the leading large $N$ contribution, but it cancels out when restricting to the connected part of the vev in the numerator of \eqref{2ptLL}. The last line, order $1/N$, is suppressed in the large $N$ limit. The remaining intermediate lines can be expressed in terms of the Green's functions studied in the previous section. After all these considerations one has
\begin{eqnarray}
\label{numLL1conn}
&&\langle {\rm tr} \left(\AU_{\!1}(0,\phi_1)\Phi_1(\phi_1)\AU_{\!1}(\phi_1,\phi_2)\Phi_1(\phi_2)\AU_{\!1}(\phi_2,2\pi)\right)
{\rm tr} \left(\PU_{\!2}(0,2\pi)\right)\rangle_{\rm conn}
\\
&& = N \Delta_{11}(\phi_2-\phi_1) \left[W(2\pi+\phi_1-\phi_2) K(\phi_2-\phi_1)
+ W(\phi_2-\phi_1) K(2\pi+\phi_1-\phi_2)\right.
\nonumber\\
&&\hspace{3cm}\left. +W(2\pi) W(\phi_2-\phi_1,2\pi+\phi_1-\phi_2)
\right].
\nonumber
\end{eqnarray}

Thus
\begin{align}
\langle\!\langle {\cal O}_1^L(\phi_1) {\cal O}^L_2(\phi_2)\rangle\!\rangle_{\rm ladder} = &
\frac{N \Delta_{11}(\phi_2-\phi_1)W(\phi_2-\phi_1)K(2\pi-\phi_2+\phi_1)}{K(2\pi)}
\nonumber\\
+&
\frac{N \Delta_{11}(\phi_2-\phi_1)W(2\pi-\phi_2+\phi_1)K(\phi_2-\phi_1)}{K(2\pi)}
\nonumber\\
+&
\frac{N \Delta_{11}(\phi_2-\phi_1)W(2\pi)
W(\phi_2-\phi_1,2\pi+\phi_1-\phi_2)}{K(2\pi)}.
\label{LLladder}
\end{align}

This is difficult to evaluate for general values of $\phi_2-\phi_1$ and the coupling.  In the limit $\phi_2\to\phi_1$ the exact correlator approaches that of a conformal field with conformal dimension $\Delta=1$. So does our ladder approximation \eqref{LLladder}, which approaches
$N \Delta_{11}(\phi_2-\phi_1)$ in the limit $\phi_2\to\phi_1$.

\subsubsection{Insertions in different loops}
For the case of insertions in different loops \eqref{2ptLR}, we need to evaluate the following expression 
\begin{eqnarray}
&&\langle {\rm tr} \left(\AU_{\!1}(0,\phi_1)\Phi_1(\phi_1)\AU_{\!1}(\phi_1,2\pi)\right)
{\rm tr} \left(\PU_{\!2}(0,\phi_2)\Phi_1(\phi_2)\PU_{\!2}(\phi_2,2\pi)\right)\rangle
\nonumber
\\
&& = \Delta_{12}(\phi_2-\phi_1) \langle {\rm tr} \left(
\AU_{\!1}(0,\phi_1)
\PU_{\!2}(\phi_2,2\pi)(\PU_{\!2}(0,\phi_2)\AU_{\!1}(\phi_1,2\pi)\right)\rangle
\nonumber
\\
&& = \Delta_{12}(\phi_2-\phi_1) \langle {\rm tr} \left(
\AU_{\!1}(\phi_1,\phi_1+2\pi)
\PU_{\!2}(\phi_2,\phi_2+2\pi)\right)\rangle
\nonumber
\\
&& = \Delta_{12}(\phi_2-\phi_1) \langle {\rm tr} \left(
\AU_{\!1}(0,2\pi)
\PU_{\!2}(\phi_2-\phi_1,\phi_2-\phi_1+2\pi)\right)\rangle
\nonumber
\\
&&= N \Delta_{12}(\phi_2-\phi_1)\Gamma(2\pi,2\pi|\phi_2-\phi_1),
\end{eqnarray}
with $\Delta_{12}$ the scalar propagator
\begin{equation}
\Delta_{12}(\phi) = \frac{g/N}{\alpha^{-1} -\cos \phi}.
\end{equation}
Thus
\begin{equation}
\langle\!\langle {\cal O}_1^L(\phi_1) {\cal O}^R_2(\phi_2)\rangle\!\rangle_{\rm ladder} =
\frac{N \Delta_{12}(\phi_2-\phi_1)\Gamma(2\pi,2\pi|\phi_2-\phi_1)}{K(2\pi)}. 
\end{equation}

\subsection{Large $\cos\gamma$ limit}
\label{largecos}
The limit $\cos\gamma\to\infty$ is a parametric limit for which both, the Wilson loop configuration and the dual world-sheet, have to be analytically continued. A key motivation to consider this limit is the fact that the ladder diagrams become the leading contribution \cite{Correa:2018pfn}. The reason for that is the following:
at every perturbative order ladder diagrams have additional factors of $\cos\gamma$ and therefore dominate over diagrams with vertices.
Moreover, as the large $\lambda$ limit of the ladder resummation can be computed, this provides the opportunity to test our world-sheet computations by an explicit comparison.

In the limit $\cos\gamma\to\infty$ not all the ladder diagrams are on the same footing. Diagrams in which all the propagators connect different loops dominate over the rest. Therefore, we have to set $W(\phi) =  1$ in all our Dyson equations. In particular, \eqref{improvedDysonG} becomes
\begin{equation}\label{improvedDysonGlimit}
 \Gamma (\phi_1,\phi_2|\varphi) =1+\int_{0}^{\phi_1}d\phi'\int_{0}^{\phi_2}d\phi''\,
G(\varphi +\phi''-\phi')\Gamma (\phi',\phi''|\varphi ),
\end{equation}
where now
\begin{equation}\label{G(theta)limit}
 G(\phi )\simeq g \,\frac{\cos\gamma  }{\alpha^{-1}-\cos \phi }.
\end{equation}
At this point  we apply a manipulation described in \cite{Erickson:1999qv} that allows us to find an intuitively clear solution to the above equation. We can obtain a differential equation by differentiating  \eqref{improvedDysonGlimit} with respect to $\phi_1$ and $\phi_2$
\begin{equation}
\label{diffeqGamma}
    \partial_{\phi_1}\partial_{\phi_2} \Gamma (\phi_1,\phi_2|\varphi) = G(\varphi+\phi_2-\phi_1)\Gamma (\phi_1,\phi_2|\varphi)\,.
\end{equation}
Changing coordinates
\begin{equation}
    x = \phi_1-\phi_2\,,\quad y = \phi_1+\phi_2\,, 
    \quad\Rightarrow\quad
    \partial_{\phi_1} = \partial_x+\partial_y\,,
    \quad
    \partial_{\phi_2} = \partial_y-\partial_x\,,
\end{equation}
so that \eqref{diffeqGamma} becomes
\begin{equation}
    (\partial_y^2 - \partial_x^2)\Gamma(x,y) = G(\varphi - x) \Gamma(x,y) \,.
\end{equation}
We can solve this equation with
\begin{equation}
\label{ansatzgamma}
    \Gamma(x,y) = \sum_n  \psi_n(x) e^{y\Omega_n}\,,
\end{equation}
where
\begin{equation}
    (\Omega_n^2 - \partial_x^2)\psi_n(x) = G(\varphi - x) \psi_n(x) \,.
\end{equation}
This is  a sort of Schr\"odinger problem
\begin{equation}
- \psi_n''(x) - \frac{g\cos\gamma}{\alpha^{-1} -\cos(\varphi-x)}\psi_n(x) = -\Omega_n^2 \psi_n(x)\,.
\end{equation}
which we do not need to solve exactly. In the limit we are interested in, this is the equation for a particle trapped in a very deep well. Thus, the sum \eqref{ansatzgamma} is dominated by the ground state eigenvalue, which is approximately given by the the depth of the well.
\begin{equation}
    \Omega_0^2 \simeq  \frac{g\cos\gamma}{\alpha^{-1} -1}\,.
\end{equation}
Therefore,
\begin{equation}
    \Gamma(\phi,\phi|\varphi) \simeq \psi_0(0) e^{\sqrt{\frac{\alpha g\cos\gamma}{1-\alpha}}2\phi}\,.
\end{equation}
The limit $\phi_2=\phi_1\to 0$ sets a boundary condition for $\Gamma(\phi_1,\phi_2|\varphi)$, which requires that $\psi_0(0) =1$. So, finally we have 
\begin{equation}
    \Gamma(2\pi,2\pi|\varphi) \simeq  e^{4\pi\sqrt{\frac{\alpha g\cos\gamma}{1-\alpha}}}\,.
\end{equation}
Using that in the large $\cos\gamma$ limit \cite{Correa:2018pfn}
\begin{equation}
    K(2\pi) \simeq e^{4\pi\sqrt{\frac{\alpha g\cos\gamma}{1-\alpha}}}\,,
\end{equation}
the ladder contribution to the correlator of insertions in different loops becomes
\begin{eqnarray}
\langle\!\langle {\cal O}_1^L(\phi_1) {\cal O}^R_2(\phi_2)\rangle\!\rangle_{\rm ladder} &\simeq&
N \Delta_{12}(\phi_2-\phi_1)\,.
\label{largecosresult}
\end{eqnarray}

Since in the large $\cos\gamma$ limit we expect that the ladder contribution is the leading one, it would be interesting to compare \eqref{largecosresult} with a dual string theory computation. The dual world-sheet configuration should also be analytically continued to be considered in a large $\cos\gamma$ limit. From \eqref{bg:modulus} and \eqref{bg:dtheta.final} it is possible to see that, for $0\leq s\leq 1$ and $t\gg 1$, the value of $\gamma$ becomes imaginary and very large. More precisely, $\gamma\simeq i\log\left(\tfrac{16 t}{1-s}\right)$. Eq. \eqref{bg:Kst} implies that, in this regime, the constant of motion $K$ has to be taken imaginary and large. Unfortunately, we do not have at the moment an explicit expression of the correlator $\langle\mathcal{O}^{L}(\phi)\mathcal{O}^{R}(0)\rangle$ for generic values of $K$. In order to make the comparison with \eqref{largecosresult} possible, one would need to sum the expression \eqref{LRgenericK} and then analytically continue the result for values of $K$ large and imaginary.

\section{Conclusions}\label{Sec:Conclusions}

In this manuscript we have considered inserting local operators in the correlator of two Wilson-Maldacena loops.  On the holographic side we have presented a complete account of the string world-sheet fluctuations, including the fermionic sector,  in section \ref{Sec:StringFluctuations}. This is a first step for precision holography explorations of AdS$_2$/dCFT$_1$ in non-supersymmetric, non-conformal setups.  We found a structure of bosonic fluctuations of the form $4+2+2$ and checked that it reduces, in the appropriate limits, to the familiar $5+3$ structure dictated by the ultrashort representation of the $OSp(4^*|4)$ supergroup governing the $\frac{1}{2}$-BPS configuration. It is worth remarking that despite being non-supersymmetric and non-conformal,  the configuration considered in this manuscript displays a fairly constrained structure and, at every step, we are able to track various parallels with the supersymmetric and conformal limits. In particular, we found that of the original $\frac{1}{2}$-BPS spectrum, four modes remain massless and two modes remain with $m^2L^2=2$ furnishing a deformation of the displacement multiplet. Following the holographic dictionary, we were able to analytically compute the two-point functions for these excitations corresponding to operator insertions in the same and in different loops at strong coupling. In the case of massless fields we found a closed analytic form in terms of elliptic theta functions and verified that the correlators satisfy a number of expected properties. For the massive fields we also obtained closed forms for the holographic correlators. 
   
We have also started a direct field-theoretic exploration of operator insertions in a system of two Wilson-Maldacena loops. On the field theory side we have concentrated on insertions of the scalars, $\Phi_{I=1,2,3,4}$ that do not enter in the definition of the Wilson-Maldacena loops, that is, not $\Phi_{5,6}$.  In section \ref{Sec:FTside}, we considered operators inserted in the same and in different Wilson-Maldacena loops, we obtained explicit expressions for the two-point correlators in the ladder approximation. Interestingly, the answer can be formulated using objects that were introduced in the context of the simpler case of correlators of two Wilson-Maldacena loops. 
    
There are a number of very interesting directions that our work stimulates. A natural one pertains  to pushing our analysis to the four-point correlators, extending the impressive work reported in  \cite{Giombi:2017cqn} for the supersymmetric $\frac{1}{2}$-BPS Wilson-Maldacena loop and in \cite{Beccaria:2019dws} for the non-supersymmetric Wilson loop. Note that in both  those cases conformal invariance along the defect is preserved; our situation requires taking one extra step into a non-conformal situation. In this work we did not develop the fermionic sector beyond obtaining the action for its quadratic fluctuations. Considering fermionic insertions, however, is a very interesting direction given the prospects of this fermionic sector sharing some features with the Sachdev-Ye-Kitaev model; we hope to report on these directions in the future.

Having described all the string fluctuations we have the ground work to tackle one-loop corrections to the effective action for the connected correlators of the two Wilson-Maldacena loops, plausibly setting up a precision holographic comparison not directly constrained by supersymmetry or conformal invariance. It should be emphasized  that this precision comparison will require some important advances on the field theory side. Our field theory results here, however, provide some evidence that such results could be achieved at least in certain limits. 
   
Another interesting direction would be to consider other representations for the Wilson loops. Indeed, the analysis of the totally symmetric and totally antisymmetric Wilson-Maldacena loops have produced powerful results in the context of the AdS/CFT correspondence \cite{Drukker:2005kx,Yamaguchi:2006tq,Hartnoll:2006is,faraggi:2011bb,Faraggi:2011ge,Buchbinder:2014nia,Faraggi:2014tna}. Finally, it would be interesting to explore to what extend our setup connects with problems in condensed matter physics, such as the problem of two Kondo impurities.

Beyond our technical progress in the treatment of Wilson-Maldacena loops in the AdS/CFT correspondence, there is a potential connection that would be interesting to explore further. As noted in the introduction, there is a striking similarity between the Gross-Ooguri phase transition of our set up and the transition in the Page curve for certain models of two-dimensional gravity \cite{Almheiri:2020cfm,Pasquarella:2022ibb}. Given the current interest and insights obtained from gravity in AdS$_2$ in the form of the Jackiw-Teitelboim model (see reviews \cite{Sarosi:2017ykf,Mertens:2022irh}), it is relevant to precisely clarify which parts of the techniques displayed here can be applied. We note that there are significant differences between the two setups.  The Wilson-Maldacena loop provides a context for open string/Wilson loop correspondence of nongravitational AdS$_2$/CFT$_1$ type correspondence since there is no dynamical gravity in the world-sheet. The Wilson-Maldacena loop has reparametrization invariance fixed by a static gauge that leaves invariant the symmetry $SO(2, 1) \in  SO(2, 4)$. This is to be contrasted with the emergent nature of $SO(2,1)$ in the JT context with its correspondence pseudo-Goldstone mode in the boundary which is related to spontaneously broken reparametrizations. This discussion was addressed, for example, in  \cite{Gutiez:2022uof}.  More recently, the authors of  \cite{Giombi:2022pas} have clarified that the out-of-time-order correlators in the  AdS$_2$ open string/Wilson loop correspondence display, in the appropriate regime, a Lyapunov growth that saturates the chaos bound. Moreover, in the conformal gauge, there is a reparametrization mode which in some respects resembles the Schwarzian mode while leading to $SO(2,1)$ invariant boundary correlators. We hope to address some of these fascinating topics in the future.

\section*{Acknowledgments}
We thank Simone Giombi for some clarifications regarding \cite{Giombi:2022pas} and Juan Maldacena for comments. DHC and GAS are partially supported by  PICT 2020-03749, PICT 2020-03826, PIP 02229, UNLP X791, UNLP X910 and PUE084 ``B\'usqueda de nueva f\'isica''. The work of AF is supported by CONICYT FONDECYT Regular \#1201145 and ANID/ACT210100 Anillo Grant ``Holography and its applications to High Energy
Physics, Quantum Gravity and Condensed Matter Systems.''
The work of WM is partially supported by the INFN, research initiative STEFI. LPZ is partially supported by the U.S. Department of Energy under grant DE-SC0007859, he also acknowledges support from an IBM Einstein Fellowship at the Institute for Advanced Study. The five of us are grateful to  ICTP for bringing us to Trieste under various programs (associateships (DHC, AF, LPZ), Giornate Uomo (WM) and the visiting programme (GAS)) during the initial stages of this project.

\appendix

\section{Conformal transformation of a pair of loops}\label{App:ConfTra}
The configuration of two Wilson-Maldacena loops is often depicted as two parallel loops of the same radius separated by some distance. One then proceeds to study the correlator as a function of this separation \cite{Gross:1998gk}. Here we will show that a pair of parallel loops with arbitrary radii, $r_1$ and $r_2$, and separated a distance $h$, can be mapped by a conformal transformation to either a pair of parallel loops with equal radii or to a pair of concentric loops lying in the same plane. Along the lines we will emphasize physical parameter that is to be varied.

Consider $\mathbb{R}^3$ and take two coaxial rings, i.e. have their centers on the $x$-axis and lie on two parallel planes orthogonal to the $x$-axis. Without loss of generality, one can fix one of the radii to unity and place it at $x=0$, so that the initial pair of loops is parameterized by 
\begin{equation}
\label{conf:loops.initial}
	x_1^\mu = (0,\cos \phi, \sin \phi)~,\qquad x_2^\mu = (x,r\cos\phi, r\sin\phi)~.
\end{equation}
Moreover, after adopting planar coordinates in the $yz$-plane, the angle can be dropped, so that it suffices to consider the two-dimensional vectors
\begin{equation}
\label{conf:loops.initial.2d}
	x_1^\mu = (0,1)~,\qquad x_2^\mu = (x,r)~.
\end{equation}

The transformations needed to transform this pair into either a pair with $x=0$ (concentric rings) or $r=1$ (parallel rings) are a special conformal transformation (SCT),
\begin{equation}
\label{conf:sct}
	x'{}^\mu = \frac{x^\mu +b^\mu x^2}{1+2b^\mu x_\mu + b^2 x^2}~,
\end{equation}
a scale transformation,
\begin{equation}
\label{conf:scale}
	x'{}^\mu = c x^\mu~,
\end{equation}
and a translation,
\begin{equation}
\label{conf:trans}
	x'{}^\mu = x^\mu+a^\mu~.
\end{equation}
If executed in the above order, with $b^\mu=(b,0)$, $c=1+b^2$, and $a^\mu=(-b,0)$, the vectors \eqref{conf:loops.initial.2d} transform into\footnote{All dimensions are expressed in units of the radius of the first loop.}
\begin{equation}
\label{conf:loops.final.2d}
	x'{}_1^\mu = (0,1)~,\qquad x'{}_2^\mu = (x',r')~,
\end{equation}
with
\begin{equation}
\label{conf:xr.trans}
	x' = \frac{x(1-b^2)+b(x^2+r^2-1)}{(1+bx)^2+b^2r^2}~,\qquad 	r' = \frac{r(1+b^2)}{(1+bx)^2+b^2r^2}~.
\end{equation}

It is easy to check that the combination 
\begin{equation}
\label{conf:inv.comb}
	\alpha = \frac{2r}{x^2+1+r^2}~,
\end{equation}
is an invariant of the transformation \eqref{conf:xr.trans}. More generally, if one lifts the assumption of unit radius for the first ring, then the invariant \eqref{conf:inv.comb} becomes
\begin{equation}
\label{conf:inv.comb2}
	\alpha = \frac{2r_1r_2}{x^2+r_1^2+r_2^2} = \frac{2r_1r_2}{2r_1r_2 + x^2+(r_1-r_2)^2}~.
\end{equation}
From the expression on the right it is clear that $0<\alpha \leq 1$, with $\alpha=1$ for the  case of two coincident rings, $x=0$, $r_1=r_2$.

Now, to obtain a concentric rings configuration, take  
\begin{equation}
\label{conf:b1}
	b = \frac1{2x} \left[x^2+r^2-1 \pm \sqrt{(x^2+r^2-1)^2 +4x^2}\right],
\end{equation} 
for which 
\begin{equation}
\label{conf:xr1}
  x' = 0~,\qquad r'_\pm = \frac1{\alpha} \left(1 \mp \sqrt{1 - \alpha^2}\right)~.
\end{equation}
It is easy to check that the two signs give rise to equivalent configurations, because $r'_+ r'_-=1$.

Similarly, for 
\begin{equation}
\label{conf:b2}
	b = - \frac{x\pm\sqrt{r[(r-1)^2+x^2]}}{x^2+r(r-1)}~,
\end{equation} 
one finds two parallel equal radius contours separated by
\begin{equation}
\label{conf:xr2}
  x' = \mp \sqrt{\frac{2(1-\alpha)}{\alpha}}~,\qquad r' = 1~.
\end{equation}
Again, the configurations corresponding to the two possible signs are equivalent.

\section{Geometry of embeddings}
\label{embed}

In this appendix, we review the geometry of embedded manifolds following \cite{Eisenhart} and provide the general expression for the pull-back of the spinor covariant derivative. Our notation will be as follows: Space-time coordinate indices are denoted by Latin letters ($m,n,\ldots$), while Greek letters ($\alpha,\beta,\ldots$) belong to the world-volume coordinates. The corresponding flat indices are underlined. Latin indices $i,j$ are used to label the directions in the normal bundle. They are flat indices by convention ($i=\flati{i}$).

A $d$-dimensional manifold $\mathbb{M}$ embedded in a $\tilde{d}$-dimensional manifold 
$\widetilde{\mathbb{M}}$ ($d<\tilde{d}$) is locally described by considering the space-time coordinates of the embedding, $x^m$, as differentiable functions of the variables $\xi^\alpha$ ($\alpha =1\ldots d$), which are identified as world-volume coordinates. 
This implies that the tangent vectors to the embedding are given by 
\begin{equation}
\label{embed:tangents}
	x^m_\alpha(\xi) \equiv \partial_\alpha x^m(\xi)~.
\end{equation}
They provide the pull-back of any bulk tensor onto the world-volume, foremost, the induced metric 
\begin{equation}
\label{embed:g}
	g_{\alpha\beta}= x^m_\alpha x^n_\beta\, g_{mn}~. 
\end{equation}
We shall assume that $g_{\alpha\beta}$ is non-degenerate. 
For a complete local basis of space-time vectors one needs to introduce a basis that spans the vector space orthogonal to the embedding, which is also called the normal bundle. There are $d_\perp=\tilde{d}-d$ independent such vectors, $N^m_i$ ($i=1,\ldots,d_\perp$), and we will apply the convention that these vectors satisfy, together with the tangents, the orthogonality and completeness relations 
\begin{equation}
\label{embed:ortho}
	N_i^m x_\alpha^n\, g_{mn} =0~,\qquad
	N_i^m N_j^n g_{mn} = \eta_{ij}~,\qquad
	g^{\alpha\beta} x_\alpha^m x_\beta^n + \eta^{ij} N_i^m N_j^n = g^{mn}~. 
\end{equation}
We allow the metric on the normal bundle, $\eta_{ij}$, to have arbitrary signature $(d_1, d_2)$, with $d_1+d_2=d_\perp$.
In particular, because it is flat, the $N^m_i$ are nothing but ($d_\perp$ of $\tilde{d})$ vielbeins $E^m_\flati{i}$ of a space-time frame that is locally adapted to the world-volume. The freedom of choice of the normal vectors gives rise to a local $O(d_1, d_2)$ symmetry in the normal bundle. This makes it clear that there will be, in general, a gauge field related to this symmetry.

The geometric structure of the embedding is characterized, in addition to the intrinsic world-volume curvature, by the second fundamental forms, $H^i{}_{\alpha\beta}$, which describe the extrinsic curvature, and the gauge connection in the normal bundle, 
$A^{ij}{}_\alpha = -A^{ji}{}_\alpha$. They are determined by the equations of Gauss and Weingarten, 
\begin{align}
\label{embed:gauss1}
	\hat{\nabla}_\alpha x^m_\beta &\equiv \partial_\alpha x^m_\beta + \Gamma^m{}_{np} x^n_\alpha x^p_\beta -
	\Gamma^\gamma{}_{\alpha\beta} x^m_\gamma = H^i{}_{\alpha\beta} N_i^m~,\\
\label{embed:weingarten}
	\hat{\nabla}_\alpha N^m_i &\equiv \partial_\alpha N^m_i + \Gamma^m{}_{np} x^n_\alpha N^p_i 
	- A^j{}_{i\alpha} N^m_j = - H_{i\alpha}{}^\beta x^m_\beta~.
\end{align}
In practice, given the tangent vectors $x^m_\alpha$ (as functions of the world-sheet coordinates), equations \eqref{embed:gauss1} and \eqref{embed:weingarten} are used to calculate the second fundamental forms, $H^i{}_{\alpha\beta}$, and the connections in the normal bundle, $A^{ij}{}_\alpha$, respectively.
Moreover, by using the appropriate connections, we have introduced in \eqref{embed:gauss1} and \eqref{embed:weingarten} the generalized covariant derivative, $\hat{\nabla}_\alpha$, which is covariant with respect to all indices. We use the hat to distinguish it from the ordinary world-sheet covariant derivative. For example, world-sheet fluctuations are parameterized by world-sheet scalars $\chi^i$ that are charged under the normal bundle gauge field. For these, we have
\begin{equation}
\label{embed:nabla.ex}
	\hat{\nabla}_\alpha \chi^i = \nabla_\alpha \chi^i +A^i{}_{j\alpha} \chi^j~.
\end{equation}

The integrability conditions of the differential equations \eqref{embed:gauss1} and \eqref{embed:weingarten} are the equations of
Gauss, Codazzi and Ricci, which are, respectively,
\begin{align}
\label{embed:gauss2} 
	R_{mnpq} x^m_\alpha x^n_\beta x^p_\gamma x^q_\delta &= R_{\alpha\beta\gamma\delta} 
	+ H^i{}_{\alpha\delta} H_{i\beta\gamma} - H^i{}_{\alpha\gamma} H_{i\beta\delta}~,\\
\label{embed:codazzi}
	R_{mnpq} x^m_\alpha x^n_\beta N^p_i x^q_\gamma &= 
	\hat{\nabla}_\alpha H_{i\beta\gamma} - \hat{\nabla}_\beta H_{i\alpha\gamma}~,\\
\label{embed:ricci}
	R_{mnpq} x^m_\alpha x^n_\beta N^p_i N^q_j &= F_{ij\alpha\beta}
	- H_{i\alpha}{}^\gamma H_{j\gamma\beta} + H_{i\beta}{}^\gamma H_{j\gamma\alpha}~,
\end{align}
where $F_{ij\alpha\beta}$ denotes the field strength in the normal bundle,
\begin{equation}
\label{embed:F}
	F_{ij\alpha\beta} = \partial_\alpha A_{ij\beta} - \partial_\beta A_{ij\alpha} 
	+ A_{ik\alpha} A^k{}_{j\beta} - A_{ik\beta} A^k{}_{j\alpha}~.
\end{equation}

Whereas the geometric relations above suffice for the treatment of tensors, some more work is needed for spinors. In general, space-time spinors decompose into several families of world-sheet spinors, which implies the existence of new connections that implement the geometric relations between these families. In particular, we are interested in the pull-back of the bulk covariant derivative (for spinors) onto the world-volume,
\begin{equation}
\label{embed:D.alpha}
	\hat{D}_\alpha \Psi = x_\alpha^m D_m \Psi = 
	x_\alpha^m \left( \partial_m +\frac14 \omega_m{}^\flati{np} \Gamma_\flati{np} \right)\Psi~.
\end{equation}
The space-time spin connections are defined in terms of a space-time frame $e^m_\flati{m}$, 
\begin{equation}
\label{embed:omega}
	\omega_p{}^\flati{mn} = -e_q^\flati{n} \left( \partial_p e^{q\flati{m}} +\Gamma^q{}_{pn} e^{n\flati{m}} \right)~,
\end{equation}
and an analogous relation holds for the world-volume spin connections, $\omega_\alpha{}^\flati{\beta\gamma}$. 

Let us pick a frame that is locally adapted to the embedding, 
\begin{equation}
\label{embed:local.frame}
	e^m_\flati{n} = \begin{cases} 
		x^m_\alpha e^\alpha_\flati{\alpha} &\text{for $\flati{n}=\flati{\alpha}$,} \\
		N^m_i &\text{for $\flati{n}=i$.} 
	\end{cases}
\end{equation}
Then, using \eqref{embed:gauss1} and \eqref {embed:weingarten}, it is straightforward to show that
\begin{equation}
\label{embed:pb1}
	x_\alpha^m \left( \partial_m e^q_\flati{n} +\Gamma^q{}_{mp} e^p_\flati{n} \right) = \begin{cases} 
		N^q_i H^i{}_{\alpha\beta} e^\beta_\flati{\alpha} 
		+ x_\beta^q e^{\beta\flati{\beta}} \omega_{\alpha\flati{\beta\alpha}}
		& \text{for $\flati{n}=\flati{\alpha}$,}\\
		-x^q_\beta H_{i\alpha}{}^\beta  + N^q_j A^j{}_{i\alpha}  
		&\text{for $\flati{n}=i$.} 
	\end{cases}
\end{equation}
Hence, the pull-back of the space-time spin connections onto the world-volume are
\begin{equation}
\label{embed:pb.spin.conn}
	x_\alpha^m \omega_{m\flati{\alpha\beta}} = \omega_{\alpha\flati{\alpha\beta}}~,\qquad 
	x_\alpha^m \omega_{mi \flati{\alpha}} = H_{i\alpha\beta} e^\beta_\flati{\alpha}~,\qquad
	x_\alpha^m \omega_{mij} = A_{ij\alpha}~.
\end{equation}
Consequently, \eqref{embed:D.alpha} becomes
\begin{equation}
\label{embed:D}
	\hat{D}_\alpha \Psi= \left( \partial_\alpha  + \frac14 \omega_{\alpha\flati{\beta\gamma}} \Gamma^\flati{\beta\gamma} 
	+\frac12 H_{i\alpha\beta} \Gamma^i \Gamma^\beta 
	+\frac14 A_{ij\alpha} \Gamma^{ij}\right) \Psi~.
\end{equation}
Using an appropriate decomposition of the gamma matrices, the last two terms in the parentheses are interpreted as connections relating  different families of world-sheet fermions.

\bibliographystyle{JHEP}
\bibliography{WLC_v2}

\end{document}